\documentclass[letterpaper]{JHEP3}
\usepackage{amssymb}

\usepackage{epsfig}

\usepackage{amsfonts}
\usepackage{amsmath}

 \numberwithin{equation}{section}
 
\usepackage{graphicx}

\usepackage{amssymb}
\usepackage{epsfig} 
\usepackage{amsfonts}
\usepackage{graphicx}
\usepackage{amsmath}

\newcommand{\insertplot}[5]{\begin{figure}
 \hfill\hbox to 0.05in{\vbox to #5in{\vfill
 \inputplot{#1}{#4}{#5}}\hfill}
 \hfill\vspace{-.1in}
 \caption{#2}\label{#3}
 \end{figure}}
 \newcommand{\inputplot}[3]{
 \special{ps: plotfile #1}
}
\newcounter{fig}

\newcommand{\beq}{\begin{equation}}
\newcommand{\eeq}{\end{equation}}
\newcommand{\beqs}{\begin{eqnarray}}
\newcommand{\eeqs}{\end{eqnarray}}

\numberwithin{equation}{section}
\newcommand{\be}{\begin{equation}}
\newcommand{\ee}{\end{equation}}
\newcommand{\bea}{\begin{eqnarray}}
\newcommand{\eea}{\end{eqnarray}}

\newcommand{\eins}{1\hspace{-0.56ex}{\rm I}}

\usepackage{graphicx}

\abstract{ 
We construct new static, asymptotically AdS solutions where the conformal 
infinity is the product of  Minkowski spacetime $M_n$ and a sphere $S^m$.
Both globally regular, soliton-type solutions and black hole solutions are considered.
The  black holes  can be viewed as natural AdS generalizations of the 
Schwarzschild 
black branes in Kaluza-Klein theory.
The solitons provide new brane-world 
models with compact extra-dimensions.
Different from the 
Randall-Sundrum single-brane scenario, a Schwarzschild black hole on the Ricci flat part of these branes
does not lead to a naked singularity in the bulk.
 }

\preprint{ }
 
\title{New AdS solitons and brane worlds \\with compact extra-dimensions } 
  
 \author{ Burkhard Kleihaus, Jutta Kunz and Eugen Radu   
 \\
 Institut f\"ur Physik, Universit\"at Oldenburg, Postfach 2503 D-26111 Oldenburg, Germany}

\begin{document}

\section{Introduction}

In the last decade a tremendous amount of interest has been focused on asymptotically
 Anti-de Sitter (AdS) spacetimes.
This interest is mainly motivated by the proposed correspondence between physical effects
associated with gravitating fields propagating in AdS spacetime and those of a conformal
field theory (CFT) on the boundary of AdS \cite{Maldacena:1997re,Witten:1998qj}.

However, this correspondence does not impose any 
constraint on the way of approaching the boundary of 
AdS spacetime.
Depending on the choice of 
'radial' coordinate defining the family of surfaces which approach the boundary,
the slices of constant radius can have different geometries or even different topologies.
For example, as discussed in \cite{Emparan:1999pm}, the maximally symmetric (euclideanized) AdS$_d$ has 
a wide variety of possible boundary geometries, 
$e.g.$  $S^1\times S^{d-2}$, $S^1\times  R^{d-2}$, $S^1\times H^{d-2}$, $S^{d-1}$, $R^{d-1}$, $H^{d-1}$
or even $S^{m}\times H^{d-m-1}$ (with $S^{m}$, $R^{m}$ and $H^{m}$ the $m-$dimensional sphere,  Euclidean plane 
and  hyperbolic space, respectively).

Of course, the allowed possibilities increase if instead of maximally symmetric AdS 
one takes   aymptotically (locally) AdS solutions of the Einstein  equations with negative cosmological
constant $\Lambda$.
Although a variety of solutions with
nontrivial boundary structure were investigated\footnote{This includes even boundary metrics
which are not globally hyperbolic, see $e.g.$ \cite{Gauntlett:2004cm}, \cite{Astefanesei:2004kn}.}, 
the issue of constructing bulk solutions compatible
with a given boundary metric is not yet fully explored.
The main obstacle here seems to be the extreme difficulty to solve the field equations
(and thus the generic absence of exact solutions)
for more complicated boundary metrics.

Moreover, as expected, there is no unique way to construct a bulk metric for a given boundary structure.
For example, a black hole has the same leading order expansion as the AdS background, while
bubble solutions may also be relevant.
These aspects are nicely illustrated by the following example.
Given the boundary geometry $R_t\times S^1 \times R^{d-3}$, 
there are at least three solutions of the Einstein equations.
First, there is the AdS$_d$ spacetime written in Poincar\'e coordinates  
(with suitable identifications),
then a topological black hole spacetime with a Ricci flat horizon
and finally the AdS soliton.
The latter is a globally regular 
solution of the Einstein equations with $\Lambda<0$ which was found by Horowitz and Myers in \cite{Horowitz:1998ha}
and has played in important role in conceptual developments in general relativity and in AdS/CFT.
It has been conjectured by Horowitz and Myers
that for a $R_t\times S^1 \times R^{d-3}$ boundary topology,
the AdS soliton is the minimum energy (perturbatively stable)
solution of the Einstein equations (its energy is lower than that of the AdS spacetime itself).
Moreover, regarding the AdS soliton as a reference background, 
Surya, Schleich and Witt found that a phase transition
occurs between the Ricci flat AdS black hole (where at least one of the horizon coordinates
is taken to be compact)
and the thermal AdS soliton \cite{Surya:2001vj}.
However, there is no phase transition for AdS black holes with Ricci flat horizons
when the zero mass black hole is taken as the thermal background.

The main purpose of this work is to present evidence for the existence
of a class of generalizations of the AdS soliton, with the $S^1$ circle there replaced by a sphere 
$S^m$.
As in the $m=1$ case, the boundary metric has also a Ricci flat part,
which can be taken to be Minkowski spacetime in $n-$dimensions, $M_n$.
However, for $m>1$, there is no foliation of AdS spacetime leading to a boundary metric  $M_n\times S^m$, since 
this configuration is not maximally symmetric.
This leads to a rather different set of features of the $m>1$ configurations as compared to
the Horowitz-Myers soliton, in particular the absence of free parameters in the solution\footnote{This resembles
the case of some field theory solitons in a flat spacetime background, see $e.g.$
\cite{'tHooft:1974qc}, \cite{Dashen:1974ck}, \cite{Volkov:1999cc}.}.
Moreover, we argue that the $m>1$ solitons emerge as zero event horizon radius limit of some black hole solutions
with the same conformal boundary at infinity.
These black holes have a  horizon topology
$R^{n-1}\times S^m$.

In this work we examine the general properties of both solitons and black holes
with a boundary metric $M_n\times S^m$
and compute their global charges by using a counterterm prescription.
For our ansatz, the Einstein equations reduce
to ordinary differential equations 
and, although we could not find an exact solution, it is straightforward to integrate them numerically
by matching 
 the near origin/horizon expansion to their Fefferman-Graham expansion \cite{FG}.
Given the presence of a compact sphere $S^m$ in the metric, the black holes share many properties of the
well-known Schwarzschild-AdS black holes with spherical topology of the horizon, in particular
the existence of two branches of solutions with different thermodynamical properties.

These solutions would provide the gravity dual for conformal gauge theories defined
on a fixed $M_n\times S^m$ background, 
the black holes accounting for finite temperature effects. 
Apart from that,
the existence of a flat sector of the solitons' metric
suggests a possible role of these solutions in a brane world context.
In this work we demonstrate that, by applying a cutting and pasting 
procedure analogous to that used in the Randall-Sundrum (RS) scenario \cite{Randall:1999vf},
the solitons yield new brane world models.
Apart from a Ricci flat part, which in the simplest case has a Poincar\'e
symmetry, the brane metric has an 
$m-$dimensional spherical internal space of constant, fixed radius.
An interesting feature here is that, different from the original
RS case, a Schwarzschild black hole on the Ricci flat part of the brane
does not lead to any pathology in the bulk.
Of course, there is price to pay for that.
The presence of two different sectors of the brane world metric, with different topologies,
leads to an anisotropic stress tensor on the brane.
Thus, apart from a brane tension (which is present also in the RS model)
one has to  assume the existence 
of some extra matter fields, which we take to be a topological soliton 
confined on the $m-$dimensional sphere.

Our  paper is structured as follows: the AdS solitons are discussed in the next section,
where we present both analytical and numerical arguments for
their existence.  
The basic properties of a brane world model based on these solutions
are discussed in Section 3.
In Section $4$ we present 
the results obtained by numerical calculations in the case of the 
black hole solutions.   We give our conclusions and 
remarks in the final Section.

\section{New AdS solitons}

\subsection{The action and metric ansatz}
We start with the following action  in $d$-spacetime 
dimensions
\begin{eqnarray}
\label{action}
I_0=\frac{1}{16 \pi G}\int_{\mathcal{M}} d^d x \sqrt{-g}
 (R-2 \Lambda)
-\frac{1}{8 \pi G}\int_{\partial\mathcal{M}} d^{d-1} 
x\sqrt{-\gamma}K,
\end{eqnarray}
where  $G$ is the gravitational constant in $d$ dimensions and 
$\Lambda=-(d-1)(d-2)/(2 \ell^2)$ is the cosmological constant. 
Here $\mathcal{M}$ is 
a $d$-dimensional manifold with metric $g_{\mu \nu }$, $K$ is the trace of the extrinsic 
curvature $K_{ab}=-\gamma_{a}^{c}\nabla_{c}n_{b}$ of 
the boundary $\partial \mathcal{M}$ with unit normal $n^{a}$ and induced metric $\gamma_{ab}$.

The classical equations of motion are derived by setting 
the variations of the action (\ref{action}) to zero,
\beqs
\label{einstein}
R_{\mu\nu}-\frac{1}{2}R g_{\mu\nu}-\frac{(d-1)(d-2)}{2\ell^2}g_{\mu\nu}=0.
\eeqs

We are interested in solutions of the Einstein equations 
whose spacelike infinity is the product of an $n-$dimensional Ricci flat space and 
an $m-$dimensional sphere.
In the simplest case,
such solutions can be described within the 
 following metric ansatz:
\begin{eqnarray}
\label{metric} 
ds^2=\frac{dr^2}{f(r)}+a(r) d\Sigma^2_{n}+P^2(r)d\omega^2_m,
\end{eqnarray}
where $n+m+1=d$.
In (\ref{metric}), $d\Sigma^2_{n}$ denotes an arbitrary
 Ricci flat metric;
for most of this work we shall restrict ourselves to the simplest case of
a Minkowski metric $d\Sigma^2_{n}=\sum_{i,j=1}^{n-1}\delta_{i,j}d x^i dx^j-dt^2.$
Also, $d\omega^2_{m}$ is the unit metric on $S^{m}$ and
$r$ is a radial coordinate, with $r_i \leq r<\infty$.
We shall suppose that, as $r\to r_i$, the  proper area of the $S^{m}$-sphere goes to
zero, while that of the flat subspace remains nonzero, $a(r_i)>0$.
In the asymptotic region $r\to \infty$, the solutions are supossed to be 
locally asymptotically AdS.

The functions $P,a$ and $f$
are solutions of the differential equations
\begin{eqnarray}
\nonumber 
&&P''
+\frac{P'f'}{2f}
+\frac{1}{2(d-2)}
\bigg(
(d-m-2)(d-m-1)\frac{a'P'}{a}
+(2d-m-4)(m-1)\frac{P'^2}{P}
\\
\nonumber
&&
{~~~~~~}-(d-m-2)(d-m-1)\frac{ Pa'^2}{4a^2}
-\frac{(2d-m-4)(m-1)}{fP}
+\frac{2\Lambda P}{f}
\bigg)
=0,
\\
\label{ec-regular}
&&a''
+\frac{a'f'}{2f}
+\frac{1}{(d-2)}
\bigg(
-\frac{(m-1)maP'^2}{P^2}
+\frac{((d-7)d-m^2+m+10)a'^2}{4a}
\\
\nonumber
&&
{~~~~~~}
+\frac{(m-1)ma'P'}{P}
+\frac{(m-1)ma }{fP^2}
+\frac{2\Lambda a}{f}
\bigg)
=0,
\\
\nonumber 
&&
\frac{(d-m-2)(d-m-1)P^2a'^2}{4m a^2}
+
\frac{ (d-m-1)P a'P'}{a}
+
(m-1)P'^2
-\frac{(m-1)}{f}
+\frac{2\Lambda P^2}{mf}
=0,
\end{eqnarray}
where a prime denotes a derivative with respect to $r$.
Note also that the  ansatz (\ref{metric}) still has some gauge freedom
which can be used to fix one of the functions $a,f$ or $P$, the corresponding equation
becoming a constraint.

\subsection{Known solutions}
 The cases $m=0$ and $n=0$ are rather special, as well as $n=1$.
For $m=0$ 
($i.e.$ no compact directions), the solution of (\ref{einstein})
has $a(r)=f(r)=r^2/\ell^2$,  and corresponds to a maximally symmetric AdS$_{d}$ spacetime  \cite{Emparan:1999pm}.
By defining $z=-\ell \log r/\ell$, the solution  becomes AdS in horospherical coordinates,
\begin{eqnarray}
\label{Poincare} 
ds^2=dz^2+e^{-2z/\ell} d\Sigma^2_{n},
\end{eqnarray}
with a $R^{d-1}$ boundary at infinity\footnote{Note that 
(\ref{Poincare}) is basically the main form used in RS brane world models.}.
The solution for the case $n=0$  ($i.e.$ no flat directions) 
has $f(r)=r^2/\ell^2+1$, $P(r)=r$,
which is a parametrizations of the Euclidean AdS$_d$ with a $S^{d-1}$ boundary \cite{Emparan:1999pm}.
The  global AdS spacetime can be written also within the ansatz (\ref{metric})
for $n=1$ and an arbitrary $m\geq 1$, and has $a(r)=f(r)=r^2/\ell^2+1$, $P(r)=r$, the boundary metric in this case being 
$R\times S^{d-2}$.

However, the  main known solution of interest in the context of our work has  
$m=1$, $n\geq 3$ and corresponds to the Horowitz and Myers 
AdS soliton \cite{Horowitz:1998ha}.
The most convinient choice for the metric gauge here is $a(r)=r^2$, 
which results in the following solution
of the equations (\ref{ec-regular}):
\begin{eqnarray}
\label{1-1} 
 f(r)=P^2(r)=\frac{r^2}{\ell^2}-M_0(\frac{\ell}{r})^{d-3},
\end{eqnarray}
 $M_0>0$ being an  arbitrary parameter.
The range of the radial coordinate is restricted in this case to $r\geq r_i=\ell M_0^{1/(d-1)}$, where $P(r_i)=0$.
In fact, $r=r_i$ represents the origin of the coordinate system and is  a 'bolt'. 
The solution is free of conical singularities for a periodicity 
\begin{eqnarray}
\label{per1} 
\beta=\frac{4\pi \ell}{(d-1) M_0^{1/(d-1)}} 
\end{eqnarray}
 of the compact $S^1$--coordinate. 
The mass of the AdS soliton is
$ {\cal M}=-M_0\beta \ell^{d-3} V_{x}/(16\pi G)$,
which is lower than AdS itself \cite{Horowitz:1998ha} ($V_x$ is the coordinate volume of the surface parametrized by $x^i$ (with $i=1,\dots,n-1$)).
 This result has found close agreement with the negative Casimir energy
 of non-supersymmetric field theory on $S^1\times R^{n}$ \cite{Horowitz:1998ha}.
 
  Another particular case which has been already studied in the literature is $n=2$ ($i.e.$ 
  $d\Sigma^2_{2}=dx^2-dt^2$)
  and an arbitrary dimension of the sphere, $m>1$.
    Different from the cases above, no analytic solution  is available
  here and one has to integrate the equations (\ref{ec-regular}) numerically.
 However, the main emphasis in the literature was on the black hole generalizations of 
 these solitons, which
  were interpreted in 
 \cite{Copsey:2006br}, \cite{Mann:2006yi}
 as describing 
 the AdS counterparts of the black strings in the $\Lambda=0$
 Kaluza-Klein theory.
 They present a number of interesting features and various extensions
 were studied subsequently in \cite{Brihaye:2007vm}-\cite{Brihaye:2009dm}.
 
\subsection{New $m>1$ solutions: the asymptotic expansion}

The $m>1$ solutions can be thought of as higher dimensional generalizations of the  $m=1$ AdS
soliton, with the $S^1$ direction replaced by a sphere.
Unfortunately, we could not find an exact solution in this case.
However, one can analyse the 
properties of these configurations by using a combination of analytical and numerical
methods, which is enough for most purposes.

For $m>1$,  we have found it convenient to fix the metric gauge\footnote{This choice has also 
been used in the study \cite{Mann:2006yi} of the 
 $n=2$ solitons and black string solutions.} by taking 
$P(r)=r$, such that $r_i=0$ and thus the range\footnote{In Section 3 we will cut off the radial
extent by inserting a brane at some finite $r_0$.} of the radial
coordinate is $0\leq r<\infty$.
it may be interesting to note that by taking
$a=e^{2q}$, the system (\ref{ec-regular})
reduces in this case to a single first order nonlinear differential equation for  $A=q'$
(the function $f$ can be expressed in terms of $a$ and $a'$) 
\begin{eqnarray}
\label{single-eq} 
A'+\psi_3 A^3+\psi_2 A^2+\psi_1 A+\psi_0=0,
\end{eqnarray}
where 
\begin{eqnarray}
\nonumber  
&&\psi_0=\frac{2\Lambda m(m-1)}{(d-2)(m(m-1)-2\Lambda r^2)},~~
\psi_1=\frac{m}{r}+\frac{2\Lambda r(2-3m(m+1)+d(3m-1))}{(d-2)(m(m-1)-2\Lambda r^2)},
\\
&&
\psi_2=\frac{2(d-m-1)((d-2)(m-1)m+\Lambda r^2(d-3m-2))}{(d-2)(m(m-1)-2\Lambda r^2)},
\\
&&
\nonumber
\psi_3=\frac{ r (d-m-2)(d-m-1)((d-2)(m-1)-2\Lambda r^2)}{(d-2)(m(m-1)-2\Lambda r^2)},
\end{eqnarray}
whose solutions we could not find in closed form, however.

Smoothness at $r=0$
requires that $a(r),f(r)$
have there a Taylor series consisting of even powers of $r$ only, with
$a(0)>0$ and $f(0)=1$. To order  $r^4$, the small $r$ solution reads
\begin{eqnarray}
\nonumber  
a(r)&=&
a(0)
\left(
1
-\frac{2\Lambda}{(m+1)(n+n-1)}r^2
+\frac{4\Lambda^2(n-1)}{m(m+1)^2(m+3)(m+n-1)}r^4
\right)
+O(r)^6,
\\
\label{origin}
f(r)&=&
1-\frac{2\Lambda}{m(m+1)(m+n-1)}r^2
+\frac{4\Lambda^2(n-1)m}{m(m+1)^2(m+3)(m+n-1)^2}r^4+O(r)^6,
\end{eqnarray}
in terms of one parameter $a(0)$.
As we shall see, this parameter is not arbitrary, being uniquely fixed by numerics.

The solitons are asymptotically locally AdS as $r\to \infty$.
For even $d$, the solution of the Einstein equations
admits at large $r$  a power series expansion of the form:
\begin{eqnarray} 
\nonumber
a(r)&=&\frac{r^2}{\ell^2}+\sum_{k=0}^{(d-4)/2}a_k(\frac{\ell}{r})^{2k}
-M(\frac{\ell}{r})^{d-3}+O(1/r^{d-2}),
\\
\label{even-inf}
f(r)&=&\frac{r^2}{\ell^2}+\sum_{k=0}^{(d-4)/2}f_k(\frac{\ell}{r})^{2k}
-(d-m-1)M(\frac{\ell}{r})^{d-3}+O(1/r^{d-2}),
\end{eqnarray}   
with $a_k,~f_k$ are constants depending on  
the spacetime dimension and the value of $m$ only.   Specifically, we find
\begin{eqnarray}
\label{inf2}  
&&a_0=\frac{m-1}{d-3},~~
a_1=\frac{(m-1)^2(d-m-2)}{(d-2)(d-3)^2(d-5)},
\\
\nonumber
&&a_2=-\frac{(m-1)^3(d-m-2)(38+4m+d(d+2m-21)))}{3(d-2)^2(d-3)^3(d-5)(d-7)}~,
\end{eqnarray} 
and
\begin{eqnarray}
\label{inf3}  
&&f_0=\frac{(m-1)(2d-m-4)}{(d-2)(d-3)},~
f_1=\frac{ (m-1)^2(d-m-1)(d-m-2) }{(d-2)(d-3)^2(d-5)},
\\
\nonumber
&&f_2=-\frac{(m-1)^3(d-m-2)(d-m-1)(8+d(m-4)-m)}{(d-2)^2(d-3)^3(d-5)(d-7)}~,
\end{eqnarray}   
their expressions becoming more complicated for higher $k$,  
without exhibiting a general pattern.

The corresponding expansion for odd values of the spacetime 
dimension is more complicated, with $\log (r/\ell)$ terms:
\begin{eqnarray}
\label{odd-inf}
a(r)&=&\frac{r^2}{\ell^2}+\sum_{k=0}^{(d-5)/2}a_k(\frac{\ell}{r})^{2k}
+\alpha\log(\frac {r}{\ell}) (\frac{\ell}{r})^{d-3}
-M(\frac{\ell}{r})^{d-3}+O(\frac{\log r}{r^{d-1}}),
\\
\nonumber 
f(r)&=&\frac{r^2}{\ell^2}+\sum_{k=0}^{(d-5)/2}f_k(\frac{\ell}{r})^{2k}
+(d-m-1)\alpha\log (\frac {r}{\ell}) (\frac{\ell}{r})^{d-3}
-(d-m-1)(M+c_0)(\frac{\ell}{r})^{d-3}+O(\frac{\log r}{r^{d-1}}),
\end{eqnarray}   
with $a_k$, $f_k$ still given by (\ref{inf2}),
(\ref{inf3}) and $\alpha$ and $c_0$ two new constants depending on both $d$ and $m$.
One finds that $\alpha$ can be expressed in a compact form as
\begin{eqnarray}
\label{expr-alpha}
 \alpha= a_{(d-3)/2}\sum_{k>0}(d-2k-1)\delta_{d,2k+1}, 
\end{eqnarray} 
 $e.g.$ $\alpha=-(m-1)^2(m-3)/12$ for $d=5$, $\alpha= (m-1)^3(m-5)(3m-10)/1600$ for $d=7$,
and $\alpha= (m-1)^4(m-7)(2m-7)(17m-91)/1778112$ for $d=9$.
The constant $c_0$ is given by
\begin{eqnarray}
\label{inf4}
&&c_0=0\delta_{5,d}-\frac{3}{3200}\delta_{7,d}-\frac{1475}{42674688}\delta_{9,d}+\dots~~~{\rm for }~~m=2, 
\\
\nonumber
&&c_0= -\frac{1}{300}\delta_{7,d}-\frac{4}{83349}\delta_{9,d}+\dots~{\rm for }~m=3, 
~c_0= \frac{129}{175616}\delta_{9,d}-\frac{85}{21233664}\delta_{11,d}+\dots~{\rm for }~m=4. 
\end{eqnarray}

In the above relations (\ref{even-inf}), (\ref{odd-inf}),
$M$ is an unknown constant which is uniquely determined by the
requirement of bulk regularity.
The dots denote  terms that are subleading orders
in $r$ relative to the terms written above (also, no new parameter appears in these subleading terms).

Note that in principle the AdS soliton with $m=1$ can 
also be written in the gauge used above, $P(r)=r$, such that
the range of the radial coordinate  becomes $0\leq r<\infty$.
However, the corresponding expression of the functions $a$ and $f$
are quite complicated for this choice. 
One finds $e.g.$
\begin{eqnarray}
\label{special-m1}
f(r)=\frac{2r^2}{\ell^2}\frac{1-4M(\frac{\ell}{r})^4}{1+\sqrt{1-4M(\frac{\ell}{r})^4}},~~
a(r)=\frac{ r^2}{2\ell^2}\left(1+\sqrt{1-4M(\frac{\ell}{r})^4}\right),~~~{\rm for~~} d=5,
 \end{eqnarray} 
 and
 \begin{eqnarray}
\label{special-m2}
 f(r)=\frac{4}{U(r)}(r^2-\frac{3U(r)}{2\ell^2})^2,~~
 a(r)=\frac{ U(r)}{\ell^4} ,~~~{\rm for~~} d=7,
 \end{eqnarray} 
 where
  \begin{eqnarray}
\label{special-m3}
\nonumber
&&U(r)=\frac{\ell^2}{6}
\bigg(
2r^2+\frac{2^{4/3}r^4}{
\bigg(
2r^6-27M\ell^6+3\sqrt{3}\ell^3\sqrt{M(-4r^6+27M\ell^6)}
\bigg)^{1/3}}
\\
\nonumber
&&{~~~~~~~~}+2^{2/3}\bigg(2r^6-27M\ell^6+3\sqrt{3}\ell^3\sqrt{M(-4r^6+27M\ell^6)}
\bigg)^{1/3}
\bigg).
 \end{eqnarray} 
 Also, from the point of view of the  expansions for $r\to 0$ and $r\to \infty$, the
 $m=1$ AdS solitons are different.
 For example, as $r\to 0$, one finds in that case  
 \begin{eqnarray}
\label{special-m1r0}
&&f(r)=f(0)
-\frac{(d-1)(d-4)}{2\ell^2}r^2+\frac{(d-1)^2(d-2)(d-3)}{16f(0)\ell^4}r^2+\dots,
\\
&&a(r)=a(0)(1+\frac{(d-1)}{2f(0)\ell^2}r^2+\frac{(d-1)^2(d-3)}{16f(0)\ell^4}r^4)+\dots,
 \end{eqnarray} 
 instead of (\ref{origin}) (with $f(0)=(d-1)|M|^{2/(d-1)}$ and $a(0)= |M|^{2/(d-1)}$),
 and
  \begin{eqnarray}
\label{special-m1r01}
f(r)=\frac{ r^2}{\ell^2}-(d-2)M(\frac{\ell }{r})^{d-3}+O(1/r^{d+1)},~~
a(r)=\frac{ r^2}{\ell^2}-M(\frac{\ell }{r})^{d-3}+O(1/r^{d+1)},~~
 \end{eqnarray} 
at infinity (and thus $\log r$ terms are absent in the asymptotics for any $d$). 
Note also that for $m=1$ solutions, the mass parameter $M$ takes only {\it negative} values.

 \subsection{A mass computation }

The action and mass of the new AdS configurations are computed by using the boundary
counterterm prescription \cite{Balasubramanian:1999re}.  In the usual approach
($e.g.$  for the $m=1$ AdS solitons or the Schwarzschild-AdS black holes) 
 the following boundary counterterm part is added to the action (\ref{action})
\cite{Balasubramanian:1999re,Das:2000cu}:
\begin{eqnarray}
I_{\mathrm{ct}}^0 &=&\frac{1}{8\pi G}\int d^{d-1}x\sqrt{-\gamma 
}\left\{ -\frac{d-2}{\ell }-\frac{\ell \mathsf{\Theta }\left( d-4\right) 
}{2(d-3)}\mathsf{R}-\frac{\ell ^{3}\mathsf{\Theta }\left( d-6\right) 
}{2(d-3)^{2}(d-5)}\left(\mathsf{R}_{ab}\mathsf{R}^{ab}-
\frac{d-1}{4(d-2)}\mathsf{R}^{2}\right) 
\right.
\nonumber  
\\
\label{Lagrangianct} 
&&+\frac{\ell ^{5}\mathsf{\Theta }\left( d-8\right) 
}{(d-3)^{3}(d-5)(d-7)}\left( 
\frac{3d-1}{4(d-2)}\mathsf{RR}^{ab}\mathsf{R}_{ab}
-\frac{d^2-1}{16(d-2)^{2}}\mathsf{R}^{3}\right.  \nonumber \\
&&\left. -2\mathsf{R}^{ab}\mathsf{R}^{cd}\mathsf{R}_{acbd}\left. 
-\frac{d-1}{4(d-2)}\nabla _{a}\mathsf{R}\nabla ^{a}\mathsf{R}+\nabla 
^{c}\mathsf{R}^{ab}\nabla _{c}\mathsf{R}_{ab}\right) +...\right\} ,
\end{eqnarray}
where $\mathsf{R}$ and $\mathsf{R}^{ab}$ are the curvature 
and the Ricci tensor associated with the induced metric $\gamma $. 
The series truncates for any fixed dimension, with new terms
 entering at every new even value of $d$, as denoted by the 
step-function ($\mathsf{\Theta }\left( x\right) =1$
 provided $x\geq 0$, and vanishes otherwise).

However,  
given the presence for odd $d$ of $\log(r/\ell)$ terms in the 
asymptotic expansions of the metric functions 
 (with $r$ the radial coordinate), the counterterms 
(\ref{Lagrangianct}) regularise the action for 
 even dimensions only. For odd values of $d$, 
we have to add the following extra terms to (\ref{action}) \cite{Skenderis:2000in}:
\begin{eqnarray}
I_{\mathrm{ct}}^{s} &=&\frac{1}{8\pi G}\int d^{d-1}x\sqrt{-\gamma 
}\log(\frac{r}{\ell})\left\{  
\mathsf{\delta }_{d,5}\frac{\ell^3 
}{8}(\frac{1}{3}\mathsf{R}^2-\mathsf{R}_{ab}\mathsf{R}^{ab}
)\right.
\nonumber  
\\
&&-\frac{\ell 
^{5}}{128}\left(\mathsf{RR}^{ab}\mathsf{R}_{ab}
-\frac{3}{25}\mathsf{R}^{3} 
-2\mathsf{R}^{ab}\mathsf{R}^{cd}\mathsf{R}_{acbd}\left. 
-\frac{1}{10}\mathsf{R}^{ab}\nabla _{a}\nabla 
_{b}\mathsf{R}+\mathsf{R}^{ab}\Box \mathsf{R}_{ab}
-\frac{1}{10}\mathsf{R}\Box 
\mathsf{R}\right)\delta_{d,7} +\dots
\right\}.\nonumber
\end{eqnarray}%
Using these counterterms in odd and even dimensions, 
one can construct a divergence-free boundary stress 
tensor from the total action $I=I_0+I_{\mathrm{ct}}^0+
I_{\mathrm{ct}}^s$ by defining a boundary stress-tensor: 
\begin{eqnarray}
\label{bst}
T_{ab}=\frac{2}{\sqrt{-\gamma}}\frac{\delta I}{\delta \gamma^{ab}}, 
\end{eqnarray} 
whose explicit expression for $d\leq 9$ is given in ref. \cite{Das:2000cu}.
Thus a conserved charge 
\begin{equation}
{\frak Q}_{\xi }=\oint_{\Sigma }d^{d-2}S^{a}~\xi ^{b}T_{ab},
\label{charge}
\end{equation}%
can be associated with a closed surface $\Sigma $ (with normal 
$n^{a}$), provided the boundary geometry has an isometry generated by a 
Killing vector $\xi ^{a}$ \cite{Booth}.

The conserved mass/energy ${\cal M}$ is the charge associated 
with the time translation symmetry, with $\xi =\partial /\partial t$. 
  A straightforward computation leads to the following expressions for the mass of the new AdS solitons:
\begin{eqnarray}
\label{mass-soliton}
{\cal M}=\frac{\ell^{m-1}m \Omega_m V_x}{16\pi G} (M +M_c^{(m,d)}),
\end{eqnarray} 
with  $\Omega_{m}=2\pi^{(m+1)/2}/\Gamma((m+1)/2)$ is the total area of a unit sphere in $m$-dimensions.
$M$ is the constant which enters the large $r$ expansion (\ref{even-inf}), (\ref{odd-inf}) of the solutions.
Also, $M_c^{(m,d)}$  are Casimir-like terms 
which appear for an odd spacetime dimension only,
\begin{eqnarray}
\label{Casimir-mass-soliton}
&&M_c^{(m,d)}= \delta_{m,2}\left(\frac{1}{24}\delta_{d,5}-\frac{1}{3200}\delta_{d,7}-\frac{25}{13333584}\delta_{d,9}+\dots\right)
\\
\nonumber
&&{~~~~~}+\delta_{m,3}\left(-\frac{7}{4800}\delta_{d,7}+\frac{221}{5334336}\delta_{d,9} +\dots\right)
+\delta_{m,4}\left(\frac{67}{351232}\delta_{d,9} +\dots\right) +\dots 
\end{eqnarray} 
  At least formally,
one can define a second charge of these solutions, which is the tension associated with translations
around some direction $x^k$.
Supposing some periodicity $L_k$
for this coordinate, one finds from (\ref{charge}) the tension
 \begin{eqnarray}
\label{tens-soliton}
{\cal T}_k=-\frac{\ell^{m-1}m \Omega_m V_x}{16\pi G L_k} (M +M_c^{(m,d)}),
\end{eqnarray} 
 which is fixed by the mass ${\cal M}$ of the solutions\footnote{This is consistent
with the interchange
symmetry of this type of configurations, see $e.g.$ the discussion for 
$m=1$ in \cite{Hu:2009rj}.}.
For an infinite 
$L_k$, the value of $V_x$ also diverges and one has to work with mass and tension densities 
(however, for the numerical calculations in this work, the values of $V_x$ and $L_k$
are not relevant).

Also, it is straightforward to show that,
 in the absence of an event horizon, 
 the action of a soliton is given by $I=\beta {\cal M}$,
 with $\beta$ the (arbitrary) periodicity
 of the Euclidean time coordinate.

\begin{table}[t]
\begin{center}
\begin{tabular}{|c|c|cr| }
 \hline
&  $d$  & $M$  & $a(0)$   \\
\hline
\hline           
& $5$    & $-0.1075$ & $0.6518$     \\ 
& $6$    & $-0.0205$ & $0.4703$  \\ 
& $7$    & $-0.0096$ & $0.3617$   \\ 
$m=2$
& $8$    & $-0.0019$ & $0.2907$  \\ 
& $9$    & $-0.0006$ & $0.2412$     \\
& $10$   & $-0.0001$ & $0.2050$   \\
 \hline
\hline           
& $6$     & $0.0801$ & $0.7440$    \\
& $7$     & $0.0262$ & $0.5861$   \\
$m=3$
& $8$     & $0.0083$ & $0.4801$    \\
& $9$     & $0.0026$ & $0.4071$     \\
& $10$    & $0.0007$ & $0.3485$  \\ 
 \hline
\hline           
& $7$     & $0.0439$ & $0.7970$   \\
& $8$     & $0.0106$ & $0.6590$   \\
$m=4$
& $9$     & $0.0032$ & $0.5597$   \\
& $10$    & $0.0011$ & $0.4851$    \\
 \hline  
\end{tabular}
\end{center} 
\vspace{0.5cm} 
{\small
{\bf Table 1.} 
The   values of the asymptotic mass parameter $M$ and
of the metric function $a(r)$ at the origin are shown for $m=2,3,4$ solitons
and several values of the spacetime dimension $d$ with four digits accuracy. 
}
\end{table}
\subsection{Numerical solutions}

The numerical evaluation of these solutions 
requires some care, since the constant $M$
appears as subleading term in the Fefferman-Graham expansion (\ref{even-inf}), (\ref{odd-inf}).

 The solutions of the equations (\ref{ec-regular}) were found by using two different methods.
 First, by employing a standard ordinary  differential  equation solver, we
evaluate  the  initial  conditions  (\ref{origin})
at  $r=10^{-6}$ for  global  tolerance  $10^{-12}$,
 adjusting the shooting parameter $a(0)$ and  integrating  towards  $r\to\infty$.
In practice,
the integration stops for some $r_{max}$ where the asymptotic limit 
(\ref{even-inf}), (\ref{odd-inf}) is reached with reasonable accuracy (typically we have taken 
$r_{max}\sim 20\ell$). 
In a different approach,
we have integrated the non linear ordinary differential equation  (\ref{single-eq})
with the boundary condition $A(0)=0$, by using a standard solver \cite{colsys}.
This solver involves a Newton-Raphson method for 
boundary-value ordinary
differential equations, equipped with an adaptive mesh selection procedure.
Typical mesh sizes include $10^2-10^3$ points; also we have used in this case a
compactified radial coordinate $x=r/(1+r)$, such that the domain of integration is
$0\leq x\leq 1$. 

%
\begin{figure}[ht]
\hbox to\linewidth{\hss%
	\resizebox{8cm}{6cm}{\includegraphics{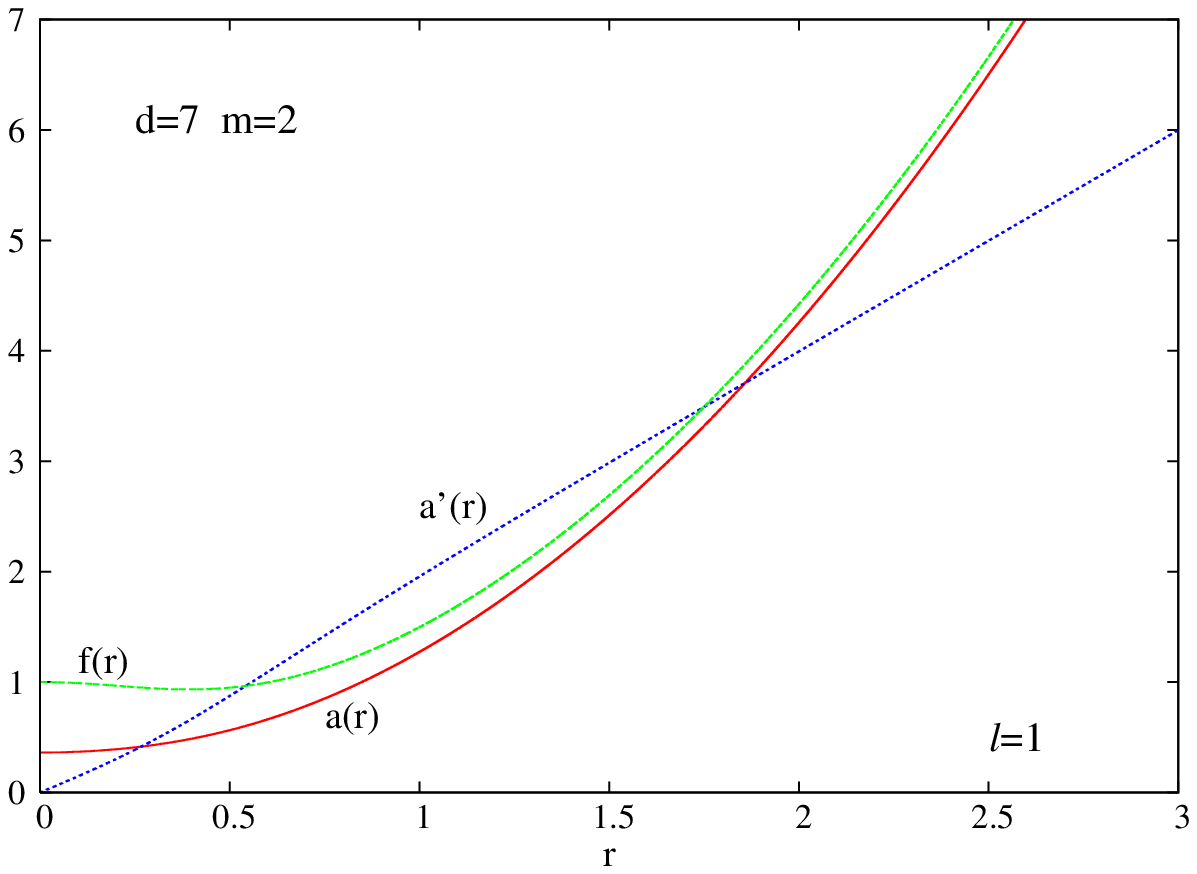}}
\hspace{5mm}%
        \resizebox{8cm}{6cm}{\includegraphics{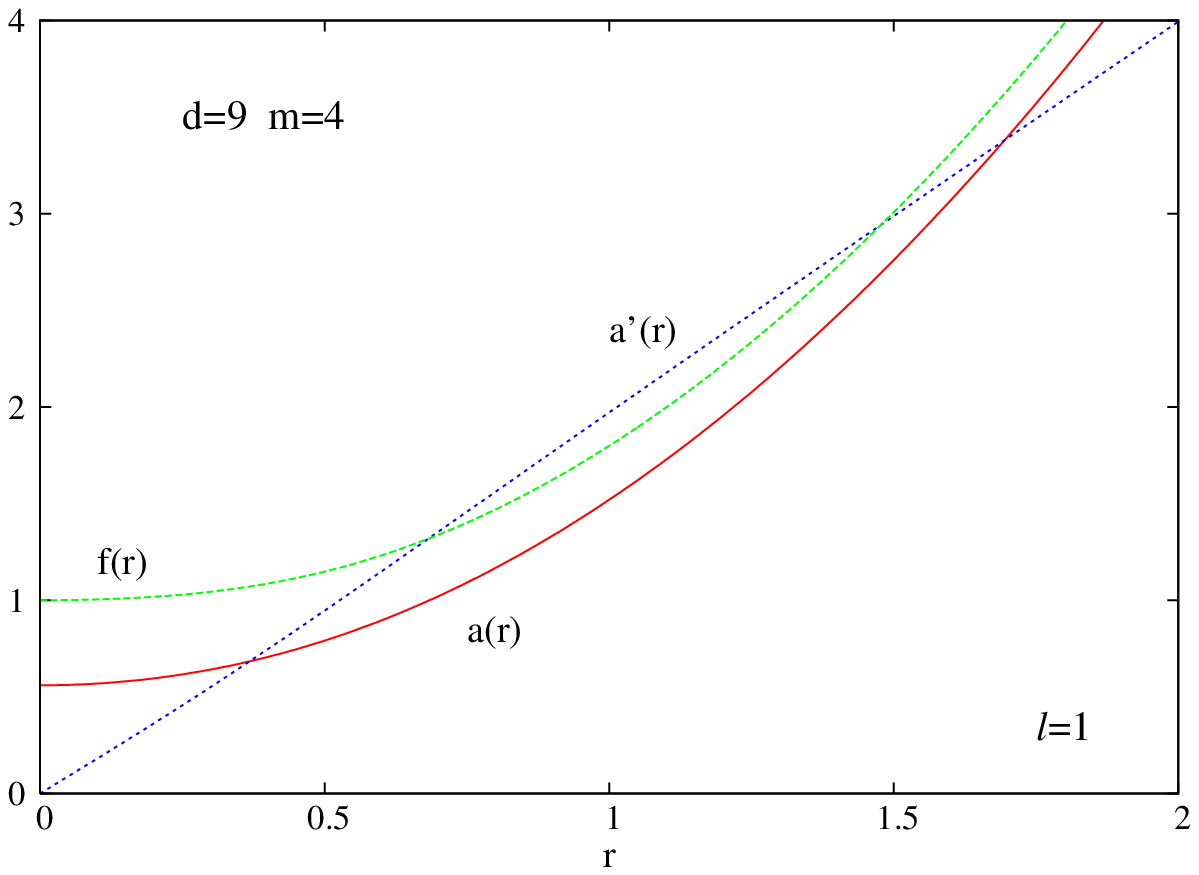}}	
\hss}

\caption{
{\small
 The profiles of the metric functions $a(r)$, $f(r)$ and 
of the derivative of $a(r)$ are shown for typical soliton solutions with $\ell=1$.}
 }
\label{fig1}
\end{figure}
 %

 AdS solutions with the asymptotics (\ref{even-inf}), (\ref{odd-inf})
 were found for $m=2,3,4,5$ and values of $d\leq 10$. 
Therefore we expect such configurations to exist for any  $(n,m)\geq 2$.  
Our numerical calculations indicate that, given 
$(d,~m,~\Lambda)$, solutions with the right asymptotics exist only for  a single value
of $a(0)$. 
This uniquely fixes also the  value of the asymptotic mass-parameter $M$ which
enters the large $r$ form of the solutions (\ref{even-inf}), (\ref{odd-inf}).
The values of $M$ and $a(0)$ are shown in Table 1 for $m=2,3,4$ solitons\footnote{The appearance of 
these values of $a(0),M$, together
with the complicated form of the $m=1$ configurations (\ref{special-m1}), (\ref{special-m2}) 
suggests that analytical 
solutions, if they exist, should be sought for another metric ansatz  than (\ref{metric}).}
 with $d\leq 10$ (note that only the first four digits are given there).
One can see that all considered solutions with $m=2$
have a negative mass-parameter $M$ (we recall that the $m=1$ solitons also have $M<0$, 
which is an arbitrary parameter).
However, this is not the case for the considered $m=3,4$ configurations.
Also, intriguingly enough, for both $m=2$ and $m=3$, the values of $\log |M|$
have almost a linear dependence on $d$.

The results in Table 1 are found for $\ell=1$. The 
solutions for any other negative values of the cosmological constant are easily 
found by using a suitable rescaling of the $\ell=1$ configurations. 
Indeed,   the Einstein 
equations 
(\ref{ep1})  are left invariant by the transformation
\begin{eqnarray}
\label{transf1} 
r \to \bar{r}= \lambda r,~~\ell \to \bar{\ell}= \lambda \ell.
\end{eqnarray}
and thus the mass of a soliton scales as ${\cal M}\to \bar {\cal M}=\lambda^{m-1}{\cal M}$.

 For all configurations we have studied,  $a(r)$ 
and $f(r)$ are smooth functions interpolating  between the corresponding values 
at $r=0$ and the asymptotic values at infinity.
We find that $a(r)$ is increasing monotonically while $f(r)$
may possess a local extremum around $r\sim \ell$.
For large $r$, the functions are proportional 
to 
$r^2/\ell^2$, indicating that the solutions
are asymptotically locally AdS as $r\to \infty$.
The profiles of the metric functions of  typical $m=2,4$ soliton solutions
are presented in Figure 1.

\section{Application: brane world models with compact extra-dimensions}

The proposal of Randall and Sundrum (RS) to localize gravity  in the vicinity
of a brane with nonvanishing tension in an AdS bulk  \cite{Randall:1999vf} has attracted enormous 
attention in the last decade.
The RS construction consists in taking two copies of a part of the five dimensional
AdS metric in Poincar\'e coordinates and gluing them
together along a boundary which is interpreted as a three brane.
In this approach, the four-dimensional gravity naturally arises at long distances on the brane
and the solution to the Einstein equations results in a single graviton zero mode (which is a
consequence of the unbroken $4d$ Poincar\'e invariance) and a continuum of Kaluza-Klein modes.

Based on the globally soliton solutions in Section 2,
 we propose in what follows a generalization
of the RS model, the brane possessing in this case an extra part 
which is a round sphere $S^m$.
The existence of $m\geq 1$ compact extradimensions is also a feature of Kaluza-Klein models.
However, the situation is quite different for the RS inspired scenario in this work.
Also, although the case of interest here is $n=4$, for the sake of generality,
we shall not fix the dimensionality of the flat part of the metric.

Note that the model with a general $m$ has some similarites with the
$m=1$ six dimensional warped brane worlds considered in various
contexts by several authors \cite{Leblond:2001xr}.
However, most the results in this Section apply also for $m=1$, in particular
the issue of a black hole on the brane. 

\subsection{The junction conditions}
 
In order to apply the RS approach to the type (\ref{metric}) of metrics,
we start by defining a new radial coordinate
\begin{eqnarray}
\label{z} 
z=\int \frac{dr}{\sqrt{f(r)}}
\end{eqnarray}
such that  
$z=r-f_2r^3/6+\dots$  as $r\to 0$   (with $f_2=\frac{4\Lambda^2(n-1)m}{m(m+1)^2(m+3)(m+n-1)^2}$).
For large $r/\ell$, one finds the usual relation $r\simeq e^{z/\ell}$.
The metric (\ref{metric}) is now written in a holographic-like form\footnote{
 In principle, the $m>1$ AdS solitons can be  directly constructed within the parametrization (\ref{metric-z}).
However, the numerics becomes more difficult in this case and we could not  
extract the asymptotic coefficient $M$
with enough accuracy.
}
\begin{eqnarray}
\label{metric-z} 
 ds^2=dz^2+a(z) d\Sigma^2_{n}+r^2(z)d\omega^2_m.
\end{eqnarray}
Now we consider the brane to be located at a given distance $z_0$
from the origin $z=0$.
The brane construction implies that we must keep the region
$0\leq z\leq z_0$
of the bulk. 
Then, by a
similar orbifold procedure as in the RS model, we replace  the part of the spacetime outside the brane $(z>z_0)$
with a copy of the inner part, ending up with a finite range for the new 'radial' 
variable\footnote{Formally, one may define a new variable $\bar z=z_0-z$ such that
the modulus of the coordinate $\bar z$
runs from the position of the brane at $\bar z=0$ and the origin of the
spacetime $|\bar z|=z_0$.} $z$.

The induced metric on the brane located at $z=z_0>0$ corresponds to a direct product
$M_n\times S^m$, but with unequal radii $R_1$ and $R_2$
for $M_n$ and $S^m$ respectively,
\begin{eqnarray}
\label{r1}
d\sigma^2= R_1^2 d\Sigma^2_{n}+R_2^2d\omega^2_m,
\end{eqnarray}
with $R_1 =\sqrt{a(z_0)}$, $R_2=r(z_0)$.
Note that only the ratio $R_2/R_1$ is relevant here,
since one can always set $R_1=1$ by using a suitable rescaling.
Therefore the value of $R_2$ can be arbitrarily small, for a position of the brane 
close to the bulk origin, $z=0$.
 
The geometry of the RS model is also given by (\ref{metric-z}), (\ref{r1}) 
with $m=0$ and $a(z)=e^{-2|z|/\ell}$.
 The range of the 'radial' coordinate  here is unbounded, $\ell\leq  z< \infty$,
 with $z\to \infty$ corresponding to a (bulk) horizon. 
Thus, from some point of view,
 the compact extradimensions on the brane affect
 also the bulk geometry and convert the AdS horizon into a regular origin.

Typically, a brane geometry is supported by some matter fields confined on the
brane.
These matter fields have an energy-momentum tensor $T_{ij}$
which enters
the Israel junction conditions on the brane
\begin{eqnarray}
\label{c1}
K_{ij}-K h_{ij}=\frac{\kappa^2}{2} (-\sigma  h_{ij}+T_{ij}),
\end{eqnarray}
where $\kappa^2=8\pi G$,
$K_{ij}$ is the extrinsic curvature tensor and  $\sigma$ is the brane tension (which 
can also be thought as a kind of matter distribution).  

The Israel junction conditions applied to a brane world (\ref{metric-z}) 
with compact dimensions
lead to the following set of equations
\begin{eqnarray}
\label{c2}
m\frac{\dot r (z_0)}{r_0}+(n-1)\frac{\dot a(z_0)}{2a_0} 
=\frac{\kappa^2}{2}(-T_{ x}^{  x}+\sigma),~~
(m-1)\frac{\dot r (z_0)}{r_0}+n\frac{\dot a(z_0)}{2a_0} 
=\frac{\kappa^2}{2}(-T_{\phi}^{\phi}+\sigma),
\end{eqnarray}
where $T_{x}^{ x}$ and $T_{\phi}^{\phi}$
are the relevant nonvanishing components of the energy-momentum tensor 
of the matter fields on the brane, with $x$ a direction on $M_n$ and $\phi$
an angle on the $m-$dimensional sphere (thus 
 $T_i^j=T_x^x\delta_{x^i}^{x^j}+T_\phi^\phi\delta_{\varphi^i}^{\varphi^j}$).
Also, $a_0=a(z_0)$, $r_0=r(z_0)$
and a dot denotes a derivative with respect to $z$ (we recall
$d/dz=\sqrt{f}d/dr$).

\subsection{Matter sources for the brane world}

From (\ref{c2}), one can see that the energy-momentum tensor 
of the matter fields on the brane is anisotropic, since
\begin{eqnarray}
\label{nc2}
\frac{\kappa^2}{2}(T_{\phi}^{\phi}-T_{x}^{x})=\frac{\dot r (z_0)}{r_0}-\frac{\dot a(z_0)}{2a_0}>0,
 \end{eqnarray}
 for any choice of $z_0$ 
(this is implied by the numerical results in Section 2).
Therefore one cannot interpret $T_{i}^{j}$ as being due to a pure tension.
 
However, one can imagine different mechanisms which lead to an energy-momentum tensor
consistent with (\ref{c2}).
Perhaps the most natural possibility is to take some matter fields
effectively living on the $S^m$ part of the brane world metric.
A simple choice here is to consider a multiplet of $(m+1)$
scalar fields $\Phi^a$, with a lagrangean density 
 (the index $i$ runs over all coordinates on the brane)
\begin{eqnarray}
\label{Lphi}
L=-\frac{1}{2}\partial_i \Phi^a \partial^i \Phi^a-V(\Phi),
\end{eqnarray}
 with some potential $V(\Phi)$.
Then we consider a hedgehog configuration for these scalars with
\begin{eqnarray}
\label{ phi}
 \Phi^a =\phi(r_0) \hat u_a
\end{eqnarray}
with $\hat u_a$ a unit vector depending on the coordinates on $S^m$ 
only\footnote{Specifically, if one takes the metric on $S^m$ as
$d\omega_m^2=d\varphi_1^2+\sin^2 \varphi_1^2(d\varphi_2^2+\dots +\sin^2 \varphi_{m-2}^2(d\varphi_{m-1}^2+
\sin^2 \varphi_{m-1}^2 d\varphi_{m}^2$)\dots), one writes $\hat u_1=\cos \varphi_1$, 
$\hat u_2=\sin \varphi_1 \cos \varphi_2,\dots,\hat u_{m+1}=\sin \varphi_1 \sin \varphi_2\dots\sin \varphi_m$.}
and $\phi(r_0)$ is the amplitude of the scalars which is constant, $\phi(r_0)=\eta$.

This leads to the simple relations
\begin{eqnarray}
\label{cond-gen}
 \kappa^2 \eta^2 =  2r_0^2 ( \frac{\dot r(z_0)}{r_0}-\frac{\dot a(z_0)}{2a_0} ) ,~~~
 \kappa^2 \sigma = \frac{m \dot r(z_0)}{r_0}+(m+2(n-1))\frac{\dot a(z_0)}{2a_0}-\kappa^2 V(\eta),
\end{eqnarray}
for the  value of the scalar field and brane tension.
Based on the expansion (\ref{origin}),
one can write an
approximate form for $\eta^2$ and $\sigma$ for small values of $z_0$ ($i.e.$ $z\to 0$)
\begin{eqnarray}
\label{sq1}
\kappa^2\sigma=\frac{m}{z_0}-\kappa^2 V(\eta)+\frac{\Lambda}{12}(m-16)z_0+O(z_0^3),~~~
\kappa^2 \eta^2=2 z_0+O(z_0^3),
 \end{eqnarray}
and, from (\ref{even-inf}), (\ref{odd-inf}), for  $z_0\to \infty$
 \begin{eqnarray}
\label{sq2}
\kappa^2\sigma=\frac{2(d-2)}{\ell}-\kappa^2 V(\eta)+\dots,~~~
\kappa^2 \eta^2= \frac{2(m-1)\ell}{(d-3)} +\dots
 \end{eqnarray}
 The scalar field equation
 \begin{eqnarray}
\label{sq3}
 \frac{m\eta}{r^2_0}=\frac{\partial V}{\partial \phi}\bigg|_{\eta},
  \end{eqnarray}
provides an extra-constraint 
which implies
that $\eta$
cannot be an extremum of the potential.
Nevertheless, a simple quadratic scalar potential, $V=\mu^2 \phi^2$,
is  compatible with (\ref{sq3}).
 
However, the choice of the matter fields on the brane proposed above 
is  not unique. 
For example,
a brane world with a compact $S^2$ can be supported by a magnetic monopole.
In this case one considers a U(1) field with 
\begin{eqnarray}
\label{Lm2s}
L=-\frac{1}{4}F_{ij}F^{ij},
\end{eqnarray}
where $F=dA$.
For an abelian magnetic monopole, the only component of the $U(1)$ field potential is  $A=Q_M\cos \varphi_1 d\varphi_2$
(with $d\omega_2^2=d\varphi_1^2+\sin^2\varphi_1 d\varphi_2^2$).
The Israel junction conditions lead to the following equations
 \begin{eqnarray}
 \kappa^2 \sigma = \frac{(2d-7)\dot a(z_0) }{2a_0}+\frac{3\dot r(z_0)}{r_0} ,~~
\frac{ \kappa^2 Q_M^2}{2r_0^4}= -\frac{\dot a(z_0)}{2a_0}+\frac{\dot r(z_0)}{r_0},
\end{eqnarray}
which fix the brane tension and the value of the magnetic charge $Q_M$  as a function of the 
position of the brane and the bulk geometry.
 
The $m>2$ brane worlds are supported by higher dimensional nonabelian generalizations of the
Dirac monopole. The Lagrangean density in this case is
\begin{eqnarray}
\label{Lm2}
L=-\frac{1}{4 g^2}{\rm Tr} \{F_{ij}F^{ij} \},
\end{eqnarray}
 $F_{kj}=\partial_k A_j-\partial_j A_k-i[A_k, A_j]$ being the gauge field 
strength tensor and $g$ the gauge coupling constant. 

 In what follows we shall use the notations and conventions of \cite{Brihaye:2002hr}.
Adopting the criterion of employing chiral
representations, both for {\it even} and for {\it odd} $m$, it is
convenient to choose the gauge group to be $SO(\bar m)$. We shall therefore
denote our representation
matrices by $SO_{\pm}(\bar m)$, where $\bar m=m+2$ and $\bar m=m+1$ for
{\it even} and {\it odd} $m$ respectively.
In this unified notation (for odd and even $m$), the spherically symmetric
Ansatz for the $SO_{\pm}(\bar m)$-valued gauge fields then reads \cite{Brihaye:2002hr}
\be
\label{YMsph}
A_i= \left(\frac{w(r_0)-1}{r_0}\right)\Sigma_{ij}^{(\pm)}\hat u_j\ , \quad
\rm{with}~~~
\Sigma_{ij}^{(\pm)}=-\frac{1}{4}\left(\frac{1\pm\Gamma_{\bar m +1}}{2}\right)
[\Gamma_i ,\Gamma_j]\ ,
\ee
where the $\Gamma$'s denote the $\bar m$-dimensional gamma matrices and
$i,~j=1,2,...,m+1$ for both cases. 
Also, $\hat u_j$ is a unit vector depending on the coordinates on $S^m$ only (see footnote 10).

One can easily verify that  $w(r_0)=0$ is a solution of the Yang-Mills equations.
Thus the nonabelian solutions are such that the field strength has components
on the $m-$sphere only
(and thus they are essentially different from the higher dimensional
nonabelian solutions  reviewed in \cite{Radu:2009rs}),
being akin to the Yang monopoles in \cite{Gibbons:2006wd}, \cite{Mazharimousavi:2008ap}.

A straightforward computation based on (\ref{c1})
 leads to the following equations
 \begin{eqnarray}
 \kappa^2 \sigma = \frac{3m}{2}\frac{\dot r(z_0)}{r_0}+(4d-3m-8)\frac{\dot a(z_0)}{4a_0},~~~
  \kappa^2 \frac{n_{\bar m}}{4 g^2} = \frac{r_0^4}{2(m-1)}(\frac{\dot r(z_0)}{r_0}-\frac{\dot a(z_0)}{2a_0}),
\end{eqnarray}
(with ${n_{\bar m}=\mbox Tr}\ \eins$, where the dimensionality
of the unit matrix is determined by the chiral representations appearing in (\ref{YMsph}))
which fix the brane tension and the gauge coupling constant
as a function of the brane's position.

For $m=4$, one can take instead a BPST nonabelian instanton \cite{Belavin:1975fg} to
support the junction conditions (\ref{c1}).
The matter lagrangean in this case is still given by (\ref{Lm2}).
The gauge
group is SU(2) with a gauge field potential 
\begin{eqnarray}
\label{BPST}
 A^a=(w(\varphi_1)+1)\,\theta^a.
\end{eqnarray}
In this relation,  $\theta^a$ are the left-invariant forms on $S^3$, with $d\omega_3^2=\delta_{a,b}\theta^a\theta^b$,
the metric on the four sphere being written as $d\omega_4^2=d\varphi_1^2+\sin^2\varphi_1 d\omega_3^2$.
This gauge field is a self-dual solution of the Yang-Mills equations on the brane for $w(r)=\cos \varphi_1$,
and thus has $T_\phi^\phi=0$. 
However, the components of the energy
momentum tensor along the flat directions $T_x^x$ are nonvanishing.

Then the junction conditions (\ref{c1}) lead to the following relations for the gauge coupling constant and 
brane tension as a function of 
 the position of  the brane:
\begin{eqnarray}
\frac{\kappa^2 \sigma}{2}= \frac{(d-5)}{2}\frac{\dot a(z_0)}{a_0}+\frac{3 \dot r(z_0)}{r_0} ,~~
 \frac{ \kappa^2}{8 g^2 r_0^4}= \frac{\dot r(z_0)}{r_0}- \frac{\dot a(z_0)}{2a_0}.
\end{eqnarray}
By using the generalizations of the BPST instanton in \cite{Tchrakian:1984gq},
one can find a similar solution for any dimension $m=4p$ of the sphere.

Other choices of possible sources to support  the junction conditions (\ref{c1}) 
seems to be possible\footnote{The possibility to employ the stress tensor
of a quantum field in a $M_n\times S^n$ geometry is especially worth investigating.
The results in \cite{Candelas:1983ae} show that such metrics  appear as
solutions of the Einstein gravity plus free 
massless fields equations in  $4+m$ dimensions.
The energy momentum tensor responsable for the curvature is produced by the quantum fluctuations
in the matter fields.}.
It would be interesting to explore the validity for $m>1$
of a number of proposals employed in models with a single extra-direction
\cite{Leblond:2001xr}.

\subsection{Black hole on a brane with compact extra-dimensions}
A curious problem of the RS model
consists in the absence of a satisfactory
solution for a black hole on the brane\footnote{ Note that there are a number of theoretical arguments against the existence 
of {\it static} black holes on the brane in the 
RS model, mainly based on a version of the conjectured
AdS/CFT correspondence \cite{Tanaka:2002rb}, \cite{Emparan:2002px}.
This seems to be confirmed by the recent numerical
results in \cite{Yoshino:2008rx}, which indicate that the static black holes on the brane in
\cite{Kudoh:2003xz}
are essentially numerical artifacts.
},
despite of a large amount of work in this direction.
The simplest proposal consists in replacing the Minkowski metric on the brane by a Schwarzschild black hole  \cite{Chamblin:1999by}.
This results in the 
bulk geometry 
\begin{eqnarray}
\label{Schw-RS} 
ds^2= dz^2+e^{-2|z|\ell}
\left(
\frac{d\rho^2}{1-(\frac{\rho_0}{\rho})^{d-4} }+\rho^2d\omega_{d-3}^2
-(1-(\frac{\rho_0}{\rho})^{d-4} )dt^2
\right) .
\end{eqnarray}
One problem with this proposal is that the $\rho=0$ singularity
extends all the way out to the AdS horizon and at this surface the solution
becomes nakedly singular\footnote{This can be seen by computing the Kretschmann scalar, which
diverges as $\rho_0^{2(d-4)}e^{4z/\ell}/\rho^{2(d-2)}$ as $z\to \infty$.} \cite{Chamblin:1999by}.
Moreover, this solution suffers from a classical Gregory-Laflamme instability \cite{Gregory:2000gf}.

Here we argue that the naked singularity is absent for 
models with compact extra-dimensions on the brane.
The simplest black hole  solution in this case is found by taking again
$d\Sigma^2_{n}$ to represent the Schwarzschild black hole  in $d-m-1$ dimensions
\begin{eqnarray}
\label{metricSc} 
ds^2= dz^2 +a(z) \left(\frac{d\rho^2}{1-(\frac{\rho_0}{\rho})^{n-3}}+\rho^2d\omega_{n-2}^2
-(1-(\frac{\rho_0}{\rho})^{n-3})dt^2\right)+r^2(z) d\omega^2_m.
\end{eqnarray}
Although  similar to the RS case,  the $\rho=0$ singularity extends out in the bulk all the  way to $z=0$,
this time the  Kretschmann scalar is finite for any value of $z$ and $\rho\neq 0$. 
For example, based on the expansion (\ref{origin}), one finds as $z\to 0$,
\begin{eqnarray}
\label{Kr} 
R_{ijkl}R^{ijkl}=f_0(\rho)+f_2(\rho)z^2+O(z^4),
\end{eqnarray}
where $f_0(\rho)$, $f_2(\rho)$ are functions of $\rho$ depending on $m,d$ and
diverging as $\rho\to 0$ only.
One finds $e.g.$ 
$f_0(\rho)=68\Lambda^2/75+12\rho_0^2/(\rho^6/a(0)^2)$,  
$f_2(\rho)=64\Lambda^3/225+16\Lambda \rho_0^2/(5a(0)^2\rho^6)$ for $m=2, d=7$
and 
$f_0(\rho)=66\Lambda^2/24+12\rho_0^2/(\rho^6/a(0)^2)$,  
$f_2(\rho)=96\Lambda^3/6125+48\Lambda \rho_0^2/(35a(0)^2\rho^6)$ for $m=4, d=9$.
The different behaviour as compared to the RS model originates in the different
properties of the bulk metrics.
In the  RS model, the AdS origin corresponds to a horizon which is  infinitely
 far from the brane  (although it can be reached by an observer
 in finite proper time).
 For the AdS solitons in this work, the horizon is absent and 
 the bulk origin is at finite proper distance from the brane.

It would be interesting to study the classical stability of these solutions.
We expect that the black holes with large enough values of $\rho_0$
as compared to $z_0$
do not possess a Gregory-Laflamme instability.

Another important problem to consider in future work is the spectrum of
linearized gravity fluctuation around the metric (\ref{metric-z}).
In the absence of an exact solution (except for $m=1$), this would
be a more difficult problem
than in the RS case.
However, we expect that the standard $1/r^{n-3}$ gravitational
potential is recovered on $M_n$ for distances
much larger than the radius of the sphere $S^m$.

\section{New black hole solutions}
The  black holes with a cosmological constant $\Lambda<0$
are of special interest in the AdS/CFT context, since they offer  the possibility  
of studying the nonperturbative structure of some CFTs. 
For example, the  Schwarzschild-AdS$_5$ Hawking-Page phase 
transition \cite{Hawking:1982dh} was interpreted as a thermal phase 
transition from a confining to a deconfining phase in the dual 
$d = 4$, ${\cal N} = 4$ super Yang-Mills theory \cite{Witten:1998zw}.

 Similar to the $\Lambda=0$ limit, the Schwarzschild-AdS  black hole solution  in $d$ dimensions
 has an event 
horizon of topology $S^{d-2}$,  
which matches the $S^{d-2}$ topology of the spacelike infinity.
However, in the presence of a negative cosmological constant, the horizon of black hole solutions
admits a much larger variety of geometries and topologies
than in the asymptotically flat case.
For example, in what follows we present arguments that 
the solitons discussed in Section 2 have black hole generalizations
with an event horizon topology
$R^{n-1}\times S^m$.
These solutions resemble
the known Schwarzschild 
black branes in Kaluza-Klein theory, since
$n-1$ flat codimensions
are present in both cases.
However, their asymptotic structure is very different, 
as well as their thermodynamical properties. 

\subsection{The equations and asymptotics}
 In the simplest case, the black hole solutions are constructed within a metric ansatz
 generalizing (\ref{metric})
\begin{eqnarray}
\label{metric-BH} 
ds^2= \frac{dr^2}{f(r)}+r^2d\omega_m^2+a(r) \sum_{i,j=1}^{n-1}\delta_{i,j}dx^idx^j-b(r)dt^2.
\end{eqnarray} 
The range of the radial coordinate is restricted here to  $r\geq r_h $
with  $b(r_h)=f(r_h)=0$, while $a(r_h)>0$. Thus $r=r_h>0$ corresponds to an event horizon.

The Einstein equations with a negative cosmological constant imply 
that the metric functions $a(r)$, $b(r)$ and $f(r)$ are solutions of 
the following equations:
\begin{eqnarray}
\label{ep2}
\nonumber 
&&a''
+(d-m-5)\frac{a'^2}{4a}
+\frac{ma'}{r}
+\frac{maf'}{(d-m-2)rf}
+\frac{a'f'}{2f}
+\frac{m(m-1)a}{(d-m-2)r^2}(1-\frac{1}{f})
+\frac{2\Lambda a}{(d-m-2)f}=0,
\\
\label{ep3} 
&&\frac{b'}{b}
+\frac{2m}{r}
+\left(  
(d-m-2)(d-m-3)\frac{r^2a'^2}{a^2}
-\frac{4m(m-a)}{f}
+\frac{8\Lambda r^2}{f}
-4m(m+1)
\right)
\\
\nonumber
&&
{~~~~~~~~~~~~~}\times\left(2r (2m+(d-m-2)\frac{ra'}{a})\right)^{-1}=0,
\\
\label{ep1} 
\nonumber
&&f'
+f\left((d-m-2)\frac{a'}{a}+\frac{b'}{b}\right)
+\frac{4\Lambda r}{d-2}
+\frac{2(m-1)}{r}(f-1)=0.
\end{eqnarray}
Unfortunately, the solutions of the above equations are known analytically only in special cases.
 For
$m=0,1$ one finds
\begin{eqnarray}
\label{ex-BH-m=1}
f(r)=b(r)=\frac{r^2}{\ell^2}-(\frac{r_0}{r})^{d-3},~~a(r)=r^2, 
 \end{eqnarray}
which corresponds to the known topological black hole with a Ricci
 flat horizon (with a compact direction for $m=1$), whose properties
 are reviewed $e.g.$ in \cite{Mann:1997iz}.
 
 For $\Lambda=0$,
 one finds instead the black brane solution in \cite{Horowitz:1991cd}, with
 \begin{eqnarray}
\label{L0Sch}
ds^2= \frac{dr^2}{1-(\frac{r_h}{r})^{d-n-2}}+r^2d\omega_m^2+ \sum_{i,j=1}^{n-1}\delta_{i,j}d x^i dx^j-(1-(\frac{r_h}{r})^{d-n-2})dt^2,
 \end{eqnarray}
  which is just the Schwarzschild black hole in $m+2$ dimensions
uplifted to $d$-dimensions.

This suggests to view the black holes discussed in this paper as natural AdS 
 generalizations\footnote{ The new AdS solitons in 
 Section 2 can also be interpreted as the AdS$_{m+1}$ regular solution with $ds^2=\frac{dr^2}{r^2/\ell^2+1}+r^2d\omega_m^2$ uplifted to $d$-dimensions
 (thus with $n$ flat extra-directions).
 This may explain why the solitons could be found only for a single value of the relevant parameters $(a(0),M)$.} 
 of the  solutions (\ref{L0Sch}), $i.e.$ as lower dimensional
 Schwarzschild-AdS black holes uplifted to $d$-dimensions.
A value $\Lambda<0$ in the bulk leads to
a nontrivial $a(r)$, to a product $g_{tt}g_{rr}\neq -1$
and also to a different asymptotic structure as compared to (\ref{L0Sch}).

 Unfortunately, there is no prescription to uplift a lower dimensional solution to higher dimensions in the
presence of a cosmological constant.
However, the  AdS counterparts of the black $(n-1)$-branes (\ref{L0Sch})  can be studied by using similar methods 
to those employed in the soliton case. 

We assume that close to $r_h$ the metric functions can be expanded\footnote{Note that the expansion (\ref{event-horizon}) 
would not hold for
extremal solutions.
However, we could not find any indication for the emergence 
of such configurations.
Nevertheless, extremal solutions are likely to exist when adding
an extra global charge to the system.}
into a Taylor series in $r-r_h$, the first terms there being:
\begin{eqnarray}
\label{event-horizon}
&&a(r)=
a_h 
\left(
1
+
\frac{2}{r_h}\frac{(d-1)\frac{r_h^2}{\ell^2}}{(d-1)\frac{r_h^2}{\ell^2}+m-1}(r-r_h)
\right)
+O(r-r_h)^2,
\\
\nonumber
&&f(r)=
\left(
(d-1)\frac{r_h^2}{\ell^2}+m-1
\right) \frac{1}{r_h} (r-r_h)
-
\left(m(m-1)+(d-1)(d-4)\frac{r_h^2}{\ell^2}\right)\frac{1}{2r_h^2} (r-r_h)^2+O(r-r_h)^3,
\\
\nonumber
&&b(r)=b_1
\left(
(r-r_h)
-\frac{ m(m-1)+(d-1)(d-4)\frac{r_h^2}{\ell^2}}{(d-1)\frac{r_h^2}{\ell^2}+m-1}\frac{1}{2r_h} (r-r_h)^2
\right)+O(r-r_h)^3,
\end{eqnarray}
in terms of two parameters  $a_h,b_1$.

Similar to the solitonic limit, the large $r$ Fefferman-Graham
expansion  of the metric functions is different for odd and even 
dimensions.
For black holes, the asymptotics is written in terms of two constants
$c_t$ and $c_z$
which appear as subleading terms in the metric functions.
For even $d$, one finds
\begin{eqnarray} 
\nonumber
a(r)&=&\frac{r^2}{\ell^2}+\sum_{k=0}^{(d-4)/2}a_k(\frac{\ell}{r})^{2k}
+c_z(\frac{\ell}{r})^{d-3}+O(1/r^{d-2}),
\\
\label{even-inf-BH}
b(r)&=&\frac{r^2}{\ell^2}+\sum_{k=0}^{(d-4)/2}a_k(\frac{\ell}{r})^{2k}
+c_t(\frac{\ell}{r})^{d-3}+O(1/r^{d-2}),
\\
\nonumber
f(r)&=&\frac{r^2}{\ell^2}+\sum_{k=0}^{(d-4)/2}f_k(\frac{\ell}{r})^{2k}
+(c_t+(d-m-2)c_z)(\frac{\ell}{r})^{d-3}+O(1/r^{d-2}).
\end{eqnarray}   
The corresponding expansion for odd values of the spacetime 
dimension is given by:
\begin{eqnarray}
\nonumber
a(r)&=&\frac{r^2}{\ell^2}+\sum_{k=0}^{(d-5)/2}a_k(\frac{\ell}{r})^{2k}
+\alpha\log(\frac {r}{\ell}) (\frac{\ell}{r})^{d-3}
+c_z(\frac{\ell}{r})^{d-3}+O(\frac{\log r}{r^{d-1}}),
\\
\label{odd-inf-BH}
b(r)&=&\frac{r^2}{\ell^2}+\sum_{k=0}^{(d-5)/2}a_k(\frac{\ell}{r})^{2k}
+\alpha\log(\frac {r}{\ell}) (\frac{\ell}{r})^{d-3}
+c_t(\frac{\ell}{r})^{d-3}+O(\frac{\log r}{r^{d-1}}),
\\
\nonumber 
f(r)&=&\frac{r^2}{\ell^2}+\sum_{k=0}^{(d-5)/2}f_k(\frac{\ell}{r})^{2k}
+(d-m-1)\alpha\log (\frac {r}{\ell}) (\frac{\ell}{r})^{d-3}
+(c_t+(d-m-2)c_z)(\frac{\ell}{r})^{d-3}+O(\frac{\log r}{r^{d-1}}),
\end{eqnarray} 
  with   $\alpha$  and $c_0$  given by (\ref{expr-alpha}) and (\ref{inf4}), respectively.
Also,
the expression of  the constants 
$a_k,~f_k$ in the above relations 
is similar to that found for the regular solutions.

For both even and odd dimensions one 
finds the asymptotic
expression of the
Riemann tensor
$R_{\mu \nu}^{~~\lambda \sigma}=-(\delta_\mu^\lambda \delta_\nu^\sigma
-\delta_\mu^\sigma \delta_\nu^\lambda)/\ell^2+\dots$, which shows that these solutions
are locally asymptotically AdS.
Note also that the soliton solutions discussed in Section 2 are recovered as $r_h\to 0$, in which case 
$c_t=c_z=-M$.

\subsection{The mass computation and a Smarr law}

The global charges of the black holes
are computed by using the same counterterm approach as in the globally regular case.
The computation of the boundary stress tensor $T_{ab}$ is straightforward and we find the 
following expressions for the mass and tension of a black hole:  
\begin{eqnarray}
\label{MT} 
&&{\cal M}=\frac{\Omega_m V_x   \ell^{m-1}}{16\pi G}\bigg[(d-m-2)c_z-(d-2)c_t\bigg]+ \frac{\Omega_m V_x   m\ell^{m-1}}{16\pi G}M_c^{(m,d)}~,
\\
&&{\cal T}_k=\frac{\Omega_m V_x  \ell^{m-1}}{16\pi G L_k}\bigg[(m+1)c_z-c_t\bigg]- \frac{\Omega_m V_x   m\ell^{m-1}}{16\pi G L_k}M_c^{(m,d)}~,
\end{eqnarray}  
with $M_c^{(m,d)}$ a Casimir term given by (\ref{Casimir-mass-soliton}) (note that the 
mass and tension are independent quantities in this case).

We note that the considered Lorentzian solutions  extremize also the 
Euclidean action as the analytical continuation $t \to i\tau$ has no 
effects at the level of the equations of motion. The Hawking 
temperature of these solutions is computed by demanding regularity of the 
Euclideanized manifold as $r \to r_h$ 
\begin{eqnarray}
T_H=\frac{1}{4\pi}\sqrt{ \left(
(d-1)\frac{r_h^2}{\ell^2}+m-1
\right) \frac{b_1}{r_h}}.
\end{eqnarray}
Thus we can proceed further by formulating gravitational 
thermodynamics via the Euclidean path integral \cite{Hawking:ig}
\[
Z=\int D\left[ g\right]  e^{-I\left[ g \right]
}\simeq e^{-I}, 
\]%
where one integrates over all metrics and matter fields between some 
given initial and final Euclidean hypersurfaces, taking $\tau $ to 
have a period $\beta=1/T_H$. Semiclassically the result is given by the 
classical action evaluated on the equations of motion, and yields to 
this order an expression for the entropy 
\begin{equation}
S=\beta M-I,  
\label{GibbsDuhem}
\end{equation}%
upon application of the quantum statistical relation to the partition 
function.
 
To evaluate the solutions' action, we integrate the Killing 
identity $\nabla^\mu\nabla_\nu  K_\mu=R_{\nu \mu}K^\mu,$
for the Killing vector $K^\mu=\delta^\mu_t$, together with the  Einstein 
equation $R_t^t={(R - 2\Lambda)/2}$. Thus, we isolate the bulk action
contribution at infinity and at $r=r_h$. The divergent 
contributions given by the surface integral term at infinity (plus the Gibbons-Hawking term)
are also 
canceled by $I_{ct}^0+I_{ct}^s$ and, together with 
(\ref{GibbsDuhem}), we find as expected $S=A_H/4G$, where
\begin{eqnarray}
\label{AH}
A_H=\Omega_{m} V_x r_h^{m} {a_h^{(d-m-2)/2}}
\end{eqnarray}
 is the event horizon area.
 
\begin{figure}[ht]
\hbox to\linewidth{\hss%
	\resizebox{8cm}{6cm}{\includegraphics{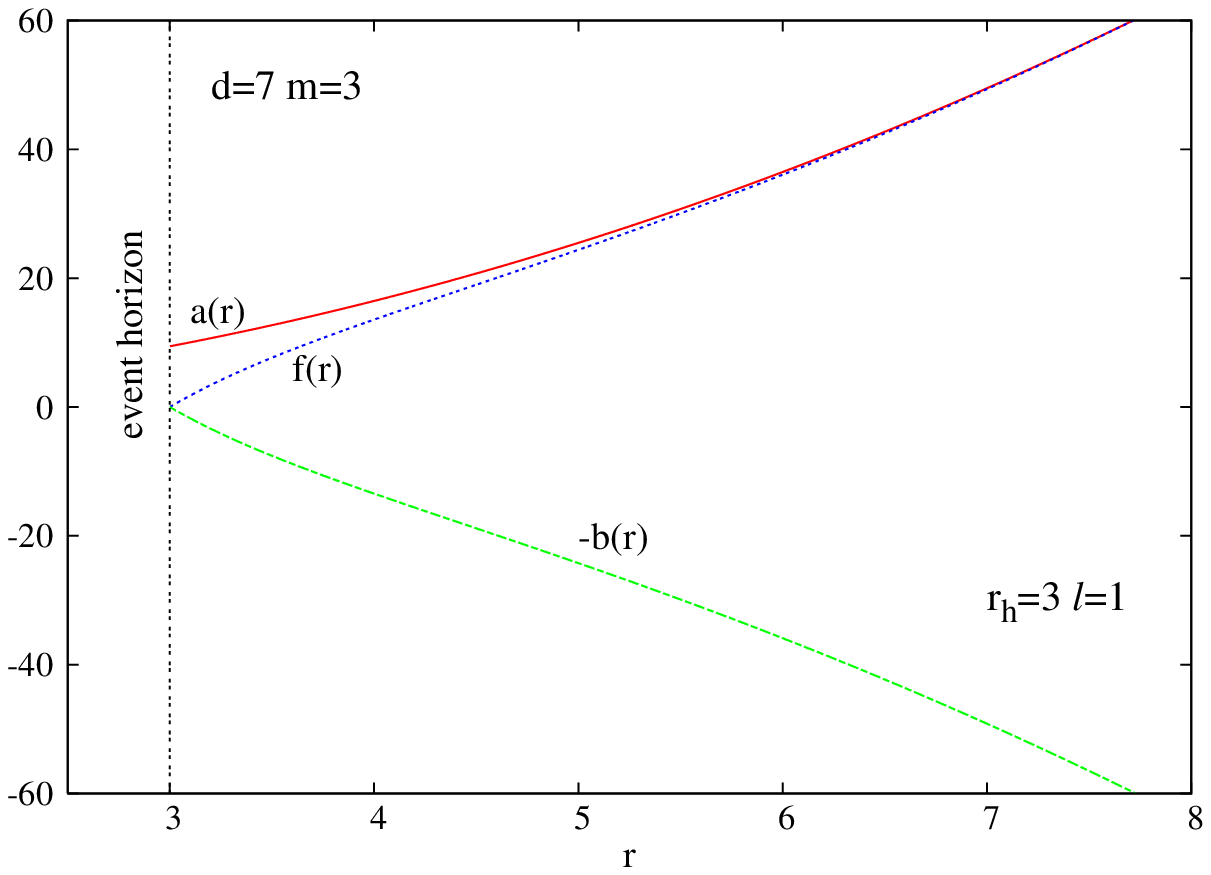}}
\hspace{5mm}%
        \resizebox{8cm}{6cm}{\includegraphics{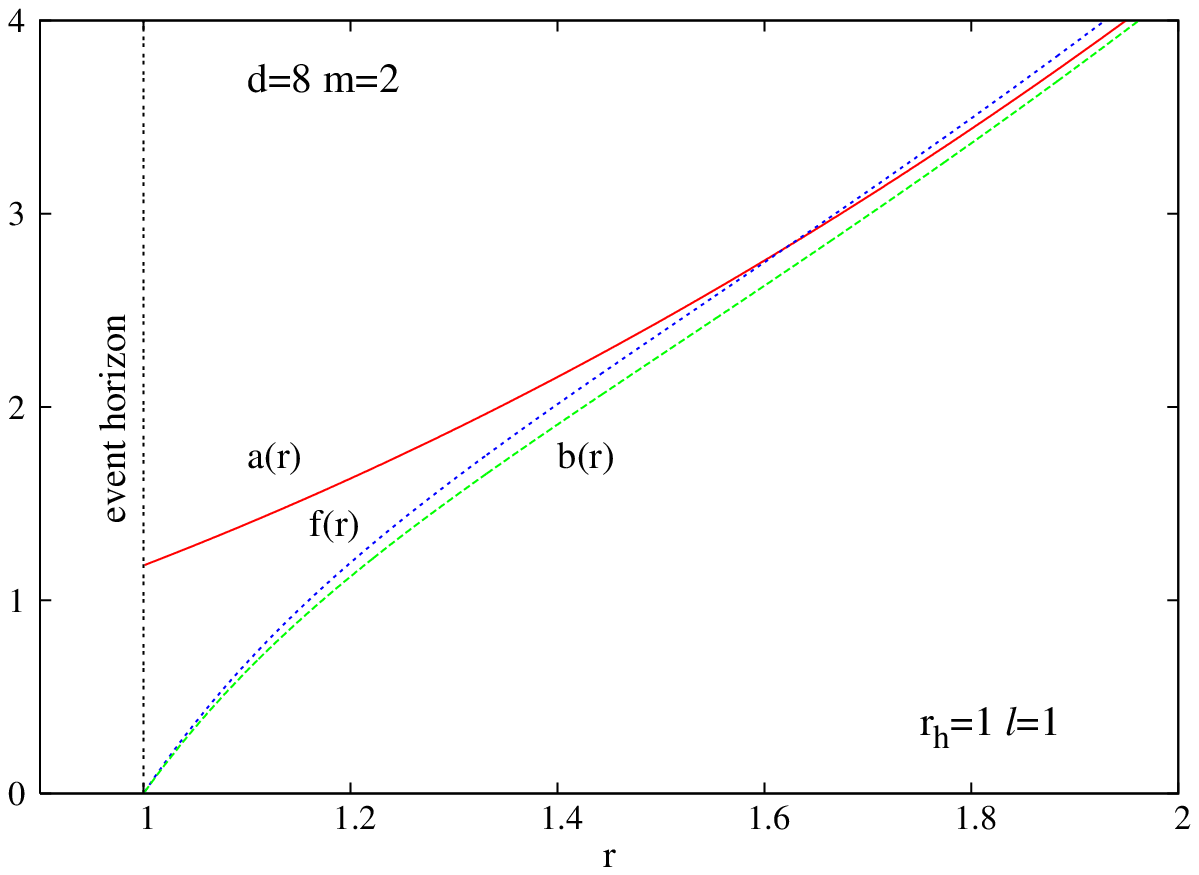}}	
\hss}

\caption{
{\small
 The profiles of the metric functions $a(r)$, $f(r)$ and 
  $f(r)$ are shown for typical black hole solutions.}
 }
\label{fig2}
\end{figure}
 
The same approach applied for a Killing vector $K^\mu=\delta^\mu_{x^k}$ yields 
the result:
\begin{eqnarray}
\label{itot2}
I =-\beta {\mathcal T}_kL_k .
\end{eqnarray}
The relations (\ref{GibbsDuhem}) and (\ref{itot2}) lead to a simple 
Smarr-type formula, relating quantities defined at 
infinity to quantities defined at the event horizon:
\begin{eqnarray}
\label{smarrform} 
M+ {\mathcal T}_k L_k =T_H S~.
\end{eqnarray} 
This relation also provides an useful check of the numerical 
accuracy (note that in all numerical data we have set $L_k=V_x=1$).

\subsection{The properties of the solutions}
We use the series expansion (\ref{event-horizon})  to fix the initial
data at $r=r_h+\epsilon$, with $\epsilon=10^{-6}$.
The system 
(\ref{ep1})-(\ref{ep3}) is then integrated by using a standard ordinary  
differential  equation solver and adjusting  for fixed shooting parameters. 
The integration stops when the asymptotic limit 
(\ref{even-inf-BH}), (\ref{odd-inf-BH}) is reached with sufficient accuracy. Given 
$(m,~d,~\Lambda,~r_h)$, solutions with the right asymptotics are found for one set of  the 
shooting parameters  $(a_h,~b_1)$ only.

\begin{figure}[ht]
\hbox to\linewidth{\hss%
	\resizebox{8cm}{6cm}{\includegraphics{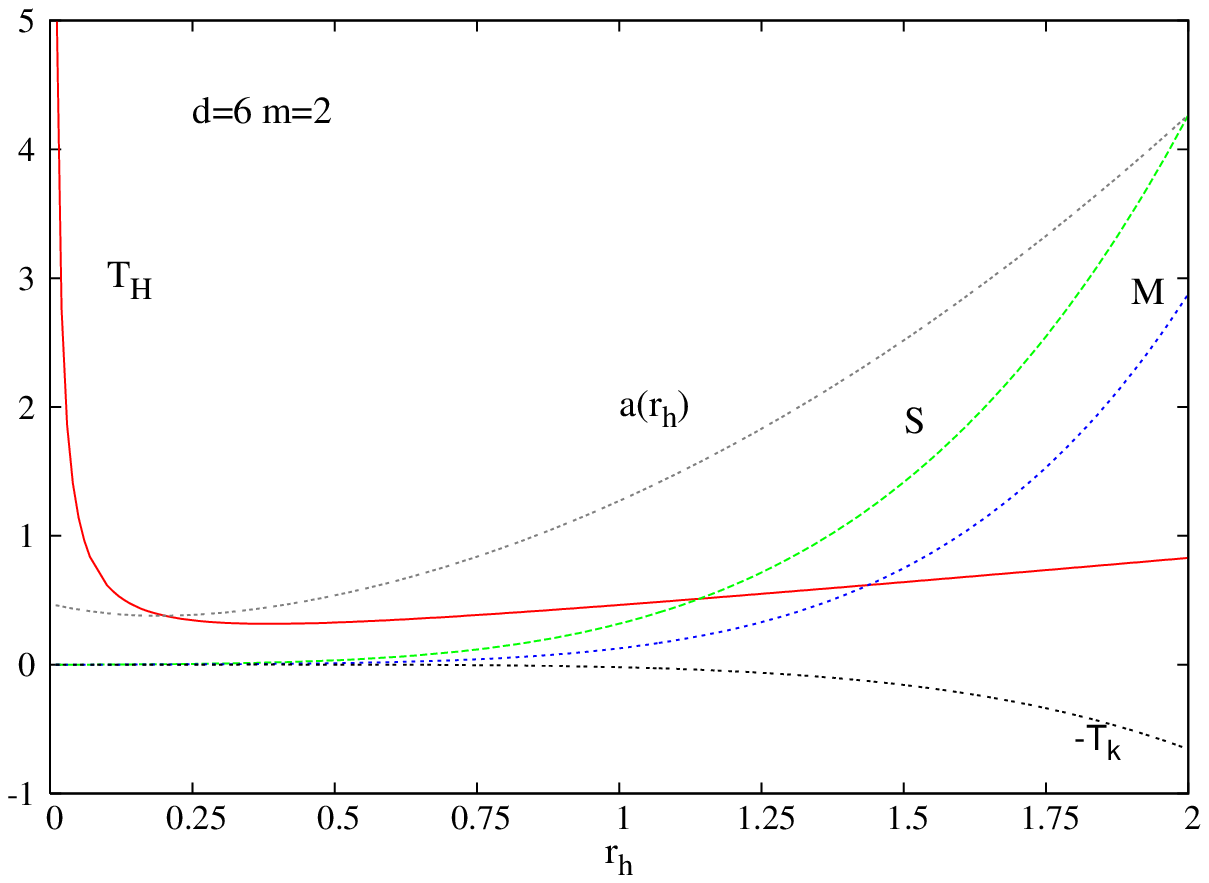}}
\hspace{5mm}%
        \resizebox{8cm}{6cm}{\includegraphics{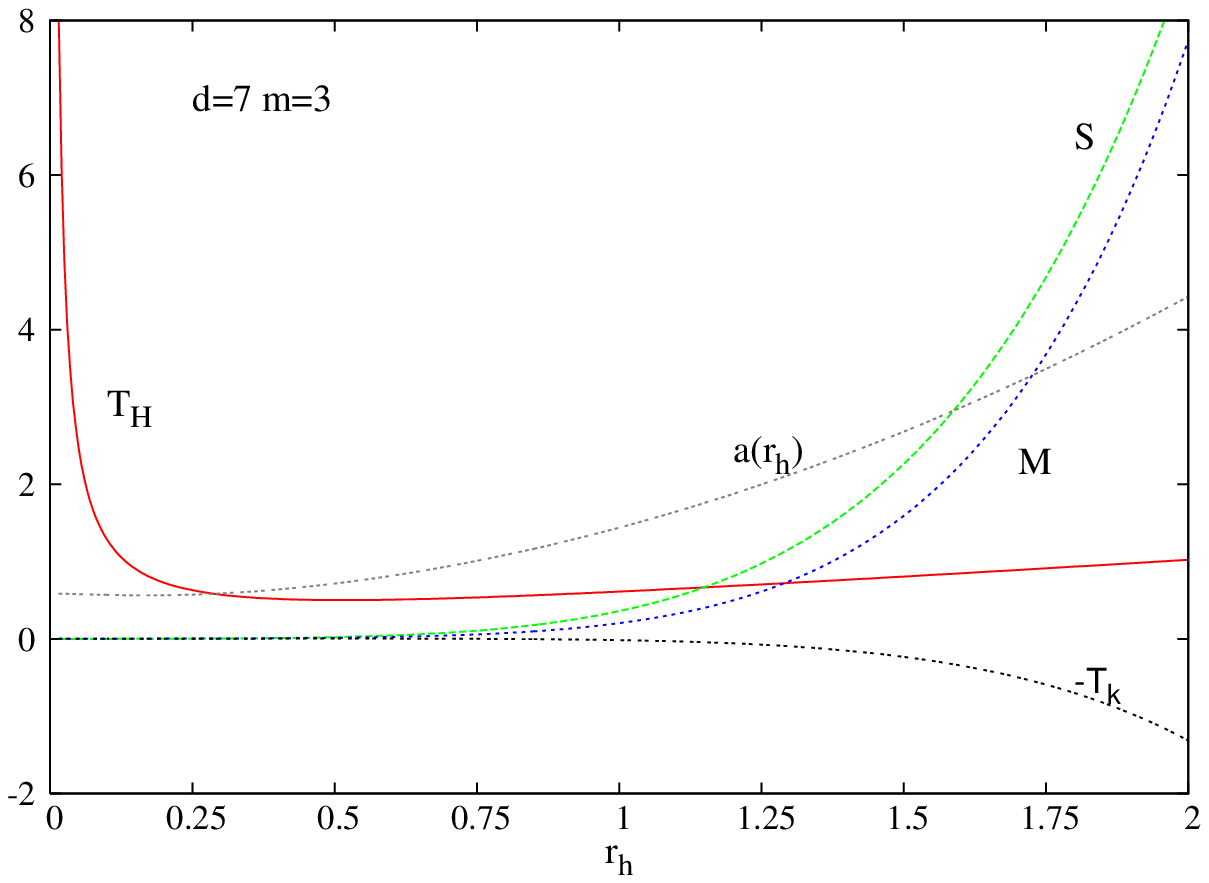}}	
\hss}

\caption{
{\small
The mass-parameter $M$, the tension 
${\mathcal T}_k$, the value  of the metric function 
$a(r)$ at the event horizon as well as the Hawking temperature 
$T_H$ and the entropy $S$ of $m=2, d=6$ and $m=3, d=7$ black hole solutions 
are represented as functions of the event horizon radius.}
 }
\label{fig3}
\end{figure}

The results we present here are obtained for $\ell=1$. However, similar to the soliton case, the 
solutions for any other value  of the cosmological constant are 
found by using a suitable rescaling of these configurations. 
The effects of the transformation (\ref{transf1}) on the  black hole solutions is
\begin{eqnarray}
\label{transf2} 
 \bar{r}_h=\lambda r_h,~~
 \bar{T}_H=T_H/\lambda ,~~
 \bar{S}=\lambda^{m-1}   S ,~~
 \bar{{\cal M}}=\lambda^{m-1} {\cal M} ,~~{\rm and}~~
 \bar{{\mathcal T}}_k=\lambda^{m-1} {\mathcal T}_k.~~~~
\end{eqnarray}
Then, given the full spectrum of solutions for $\ell=1$ (with 
$0<r_h<\infty$), one may find the corresponding branch for any value of 
$\Lambda<0$.

  We have constructed black hole solutions in all dimensions between five 
and ten with several values of $m$
and  for $0\leq r_h\lesssim 10$. 
  Thus they are  likely to exist for any allowed set $(d,~m)$ 
and for
any value of the event horizon $r_h$.

 As typical examples, the metric functions
 $a,b$ and $f$ are shown  in Figure 2 as functions of the radial
 coordinate $r$ for two values of $(d,m)$.
 One can see that the term $r^2/\ell^2$ starts dominating the profiles
 of these functions very rapidly,
 which implies a small difference between them for large enough $r$.

The dependence of various physical parameters on the event horizon 
radius is presented in Figure 3 for $m=2, d=6$ and $m=3, d=7$ solutions. 
These plots exhibit the basic features of the solutions we found also
in other dimensions and for other values of $m>1$ (note that there and in Figure 4 we set $V_x\Omega_{m}/G=L_k=1$ in the 
expressions for the mass, tension and entropy and we subtracted the constant Casimir 
 terms in odd dimensions). 

  Similarly to the spherically symmetric Schwarzschild-AdS solutions, 
one can see in Figure  $3$ that the temperature of the black holes is bounded from below. At low temperatures we have 
a single bulk solution which we conjecture to correspond to the 
thermal globally regular soliton. At high temperatures there exist two 
additional solutions that correspond to the small and large black 
holes. 
 For large black holes, the entropy is increasing with the temperature,
while the small black holes have a negative specific heat.

Moreover, the free energy $F=I/\beta$  is positive for small $r_h$ 
and negative for large $r_h$.  
This shows that the phase
transition found in \cite{Hawking:1982dh}  occurs also in this case and
there are two 
branches of solutions consisting of smaller (unstable) and large (stable) black 
holes. 
  This is illustrated in Figure
4, where the free energy is plotted versus the temperature for $d=7$ solutions with several values of $m$
(the $m=5$ configurations have $n=1$ and correspond  to Schwarzschild-AdS$_7$ black holes).
 
%
\begin{figure}[ht]
\hbox to\linewidth{\hss%
	\resizebox{9cm}{7.1cm}{\includegraphics{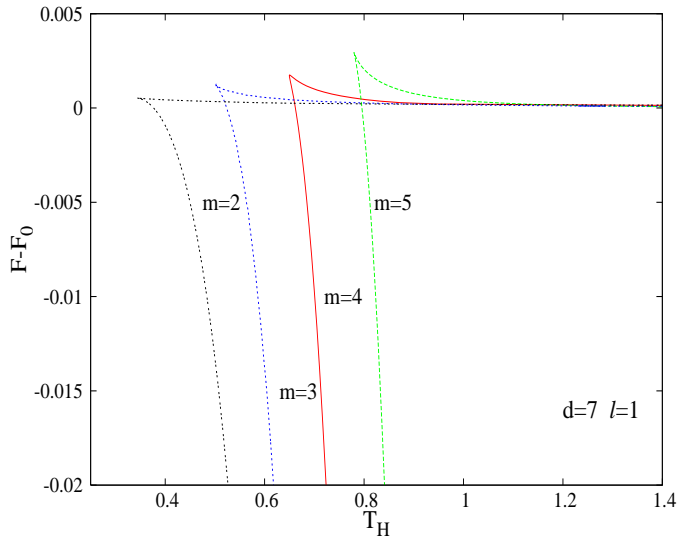}}
 \hss}
	\caption{   
	{\small
	The free energy \textit{vs.} the temperature for the small 
and large $d=7$  black hole solutions is plotted for several values of $m$. Here we have subtracted the free 
energy contribution $F_0$ of the corresponding globally regular solutions.
} 
}
\label{Fig1}
\end{figure}
 %
 
Without entering into details, we note that by
performing a double analytic continuation, the black hole solutions in this work
 describe static bubbles of nothing in AdS, with a line element:
 \begin{eqnarray}
\label{metric-b} 
ds^2=a(r)(-du^2+\sum_{i,j=1}^{n-2} \delta_{i,j}dx^i dx^j)+b(r)d\tau^2+ \frac{dr^2}{f(r)}
+r^2d\omega^2_{m},
\end{eqnarray}
where $\tau$ has a periodicity $\beta=1/T_H$.
 The properties of these solutions can be discussed in a similar way to the black hole case.
For example, there are both `small' and `large' bubbles,
 which result as analytical continuation of the corresponding black hole branches.
 Using the counterterm approach, one can show that the mass of a bubble solution is
\beqs
{\cal M}_{bubble}&=&-\beta {\cal T}_u.
\eeqs
Note also that the analytic continuation of a soliton leads to the same regular solution
(since $a(r)=b(r)$ in that case), with an arbitrary value of $\beta$.

\section{Further remarks. Conclusions}

The purpose of this work was to present evidence for the existence
of a new type of solutions of Einstein gravity with negative $\Lambda$.
For such solutions, the topological structure of the boundary at
infinity is the product
of time and $S^m\times R^{d-m-2}$, with $m>1$.
Both globally regular, soliton-type solutions and black holes have been considered.
Since we  could not find exact solutions, 
we have resorted to numerical methods.
Analytical expressions for the solutions can be constructed, however,
close to the origin $r=0$ (or to the event horizon $r=r_h>0$)
and for large values of $r$.  

The solitons were used to construct new brane world models with
compact extra dimensions.
Different from the Randall-Sundrum  \cite{Randall:1999vf}
and the Karch-Randall models \cite{Karch:2000ct},
the existence of extra-dimensions on the brane
imposes the presence of  matter fields,
which have been taken to be topological 
solitons confined on the sphere $S^m$.

It is clear that this work has  only scratched the surface of the full subject
and a variety of asymptotically (locally) AdS solutions with more complex boundary structure
are likely to be found.
For example, as in the $n=2$ case  \cite{Mann:2006yi}, 
  a generalization of the black hole solutions in this work
with the $m-$dimensional sphere  $d\omega^2_{m}$ replaced
by a hyperboloid $d\Xi^2_{m}$ should exist (note that these configuration will not possess
a soliton limit). 
In fact, it would be interesting to study a more general class of 
black hole solutions
with the line element
 \begin{eqnarray}
 ds^2=\frac{dr^2}{f(r)}+a(r) \sum_{i,j=1}^{n-1}\delta_{i,j}dx^i dx^j-b(r)dt^2+P^2(r)d\omega^2_m+c(r)d\Xi^2_{p},
\end{eqnarray}
with $m+n+p+1=d$, the solutions in our paper corresponding to $p=0$.
The metric functions $a,b,c$ and $P$ would satisfy different boundary conditions
at $r=r_h$  and thus would  not be equal.

 We did not address the question of classical stability
of the new solutions in this work.
For $n=2$, the
 results in  \cite{Brihaye:2007ju}
 show that the solutions are stable for large enough values of the event horizon radius only.
We expect that the situation will be the same for any $n\geq 2$.
This is suggested by the thermodynamical properties of the solutions,
together with the Gubser-Mitra conjecture \cite{Gubser:2000ec}
that correlates the dynamical and thermodynamical stability for systems
with translational symmetry and infinite extent. 
Therefore we expect the branch of black sole solutions with a negative specific heat to possess also
a Gregory-Laflamme unstable mode.

In connection to that, it would be of particular interest to construct AdS
black holes approaching the asymptotics (\ref{even-inf-BH}), (\ref{odd-inf-BH})
as $r\to \infty$ but with a different topology of the horizon.
For $n=2$, these would be the AdS counterparts of the $\Lambda=0$
caged black holes in Kaluza-Klein theory, see $e.g.$ Ref. \cite{Harmark:2007md}.
The existence of such configurations is suggested by the results in
\cite{Brihaye:2007ju}, \cite{Delsate:2008iv}.

Also, the configurations in this work can be used to construct new 
lower dimensional non-trivial soliton and black hole solutions of the Einstein-dilaton system
with a Liouville dilaton potential.
As with the $n=2$ case in \cite{Mann:2006yi}, these solutions are found by dimensionally reducing w.r.t.
one (or several) Killing vector(s) $\partial/\partial x^i$.
Moreover, by using the techniques in  \cite{Mann:2006yi}, one can show that the reduced action has an effective
$SL(2,R)$ symmetry, 
which can be used to add an electric charge to these lower dimensional configurations.

We close this paper with several remarks on the possible role of the solutions
in this work in the context of AdS/CFT correspondence.
The background metric upon which the dual field theory resides 
is found by taking the rescaling 
$h_{ab}=\lim_{r \rightarrow \infty} \frac{\ell^2}{r^2}\gamma_{ab}$.
Therefore, for both soliton and black hole solutions
 we find 
 \begin{eqnarray}
 \label{metric-CFT}
ds^2=h_{ab}dx^adx^b=-dt^2+\sum_{i,j=1}^{n-1}\delta_{i,j}d x^i dx^j+\ell^2d\omega_m^2,
\end{eqnarray}
and so the conformal boundary, where the dual theory lives, 
is $R_t\times R^{n-1}\times S^m$.

The expectation value $<\tau _{a}^b>$ of the stress tensor of the dual CFT 
can be computed using the  relation 
\cite{Myers:1999ps}
\begin{eqnarray} 
\sqrt{-h}h^{ab}<\tau _{bc}>=\lim_{r\rightarrow \infty }\sqrt{-\gamma 
}\gamma
^{ab}T_{bc},
\end{eqnarray}
where $T_{bc}$ is the gravity boundary stress tensor (\ref{bst}).

Let us consider for example\footnote{The expressions of $<\tau _{a}^b>$ 
for $m=2$, $d=5$ and  $m=4$, $d=7$ 
are given in \cite{Mann:2006yi}, and
\cite{Brihaye:2007vm}, respectively.} the (most interesting)
case of black holes with a four dimensional flat subspace ($i.e.$ $n=4$).
A straightforward computation gives  the following
expressions for the nonvanishing components of $<\tau^b_a>$
\begin{eqnarray}  
\label{st1}
 && <\tau^t_t>= \frac{1}{8\pi G }\frac{1+8000c_t-4800 c_z }{3200 \ell} , ~<\tau^x_x>=\frac{1}{8\pi G } \frac{1-1600 c_t+4800 c_z}{3200 \ell}, 
\\
\nonumber
&&{~~~~~~~~~~~~~~}<\tau^{\phi}_{\phi}>= \frac{1}{8\pi G }\frac{5-800c_t-2400 c_z}{1600 \ell  },
\end{eqnarray}
for $m=2$ ($i.e.$ $d=7$), and
\begin{eqnarray}  
\label{st2}
<\tau^t_t>=\frac{1}{8\pi G }\frac{6c_t-3c_z}{2\ell} , ~~<\tau^x_x>=-\frac{1}{8\pi G }\frac{c_t-4c_z}{2\ell}, 
~~<\tau^{\phi}_{\phi}>=-\frac{1}{8\pi G }\frac{c_t+3c_z}{2\ell},
\end{eqnarray}
for $m=3$ ($i.e.$ $d=8$).
The stress tensor of the dual CFT defined on an eight dimensional space with a compact $S^4$ ($i.e.$  a $d=9$ bulk) is
\begin{eqnarray}  
\nonumber
 &&<\tau^t_t>=-\frac{1}{8\pi G }\frac{221-12446784 c_t+7112448c_z}{ 3556224 \ell} ,
~<\tau^x_x>=-\frac{1}{8\pi G }\frac{221-12446784 c_t+7112448c_z}{ 3556224 \ell}, 
\\
\label{st3}
 && <\tau^{\phi}_{\phi}>=\frac{1}{8\pi G }\frac{875-5334336c_t-21337344c_z}{10668672 \ell}.
\end{eqnarray}
In the above expressions,  $<\tau^x_x>$ and $<\tau^{\phi}_{\phi}>$ stand for the 
nonvanishing components of the stress tensor of the dual CFT 
along the flat directions and on the sphere, respectively.

As expected, these stress tensors are finite and covariantly conserved.
For even $d$, we have found that $<\tau^b_a>$ is always traceless,
as expected from the absence of conformal anomalies for the boundary field theory
in odd dimensions.
However,  for  odd $d$ ($i.e.$ an even dimensional
boundary metric)
$<\tau^b_a>$  is $not$ traceless.
In fact, we have verified that for $d=7$ its trace $<\tau^{a}_a>=3/(3200\pi G \ell)$ is precisely 
equal to the conformal anomaly of the boundary CFT   in six dimensions
\cite{Skenderis:2000in}:
\beqs
 {\cal A}=-\frac{1}{8\pi G}\frac{\ell^{5}}{128}\left( {RR}^{ab}
 {R}_{ab}-\frac{3}{25} {R}^{3} 
-2 {R}^{ab} {R}^{cd} {R}_{acbd}-
\frac{1}{10} {R}^{ab}\nabla _{a}\nabla_{b}
 {R}+ {R}^{ab}\Box  {R}_{ab}-\frac{1}{10} {R}\Box 
 {R}\right),~~~{~~}
\eeqs
where $ {R}$,  $ {R}^{ab}$ and  $ {R}_{abcd}$  are the curvature 
and the Ricci and Riemann tensor associated with the  metric (\ref{metric-CFT}).  
A similar computation performed for the  case $d=5,~m=2$ leads 
to a boundary stress tensor whose trace matches precisely the 
conformal anomaly of the dual four-dimensional CFT \cite{Mann:2006yi}.
 
Further analysis of these metrics and their role in string theory remain interesting issues to explore in the future.

\section*{Acknowledgements}
B.K. gratefully acknowledges support by the DFG. 
The work of E.R. was supported by a fellowship from the Alexander von Humboldt Foundation.

 \begin{small}
 
 \end{small}
\end{document}